\documentclass[5p]{elsarticle}
\usepackage[T1]{fontenc}
\usepackage[latin9]{inputenc}
\usepackage{float}
\usepackage{multirow}
\usepackage{amsmath}
\usepackage{amssymb}
\usepackage{graphicx}

\makeatletter

\newcommand{\noun}[1]{\textsc{#1}}
\providecommand{\tabularnewline}{\\}
\floatstyle{ruled}
\newfloat{algorithm}{tbp}{loa}
\providecommand{\algorithmname}{Algorithm}
\floatname{algorithm}{\protect\algorithmname}

\makeatletter
\def\ps@pprintTitle{%
   \let\@oddhead\@empty
   \let\@evenhead\@empty
   \let\@oddfoot\@empty
   \let\@evenfoot\@oddfoot
}
\makeatother

\makeatother

\usepackage[swedish,english]{babel}
\begin{document}

\begin{frontmatter}{}

\title{Physiological Gaussian Process Priors for the Hemodynamics in fMRI
Analysis}

\author[mymainaddress]{Josef Wilzén{*}\corref{mycorrespondingauthor}}

\cortext[Josef Wilzén]{Corresponding author}

\ead{josef.wilzen@liu.se}

\author[mymainaddress,mysecondaryaddress,mythirdaddress]{Anders Eklund}

\author[mymainaddress,myfourthaddress]{Mattias Villani}

\address[mymainaddress]{Division of Statistics \& Machine Learning, Department of Computer
and Information Science, Linköping University, Linköping, Sweden}

\address[mysecondaryaddress]{Division of Medical Informatics, Department of Biomedical Engineering,
Linköping University, Linköping, Sweden}

\address[mythirdaddress]{Center for Medical Image Science and Visualization (CMIV), Linköping
University, Linköping, Sweden}

\address[myfourthaddress]{Department of Statistics, Stockholm University}
\begin{abstract}
\textbf{\noun{Background}}

\textbf{\hspace{-0.5cm}}Inference from fMRI data faces the challenge
that the hemodynamic system that relates neural activity to the observed
BOLD fMRI signal is unknown.\vspace{0.1cm}

\textbf{\hspace{-0.5cm}}\textbf{\noun{New Method}}

\textbf{\hspace{-0.5cm}}We propose a new Bayesian model for task
fMRI data with the following features: (i) joint estimation of brain
activity and the underlying hemodynamics, (ii) the hemodynamics is
modeled nonparametrically with a Gaussian process (GP) prior guided
by physiological information and (iii) the predicted BOLD is not necessarily
generated by a linear time-invariant (LTI) system. We place a GP prior
directly on the predicted BOLD response, rather than on the hemodynamic
response function as in previous literature. This allows us to incorporate
physiological information via the GP prior mean in a flexible way,
and simultaneously gives us the nonparametric flexibility of the GP.\vspace{0.1cm}

\textbf{\hspace{-0.5cm}}\textbf{\noun{Results}}

\textbf{\hspace{-0.5cm}}Results on simulated data show that the proposed
model is able to discriminate between active and non-active voxels
also when the GP prior deviates from the true hemodynamics. Our model
finds time varying dynamics when applied to real fMRI data.\vspace{0.1cm}

\textbf{\hspace{-0.5cm}}\textbf{\noun{Comparison with Existing Method(s)}}

\textbf{\hspace{-0.5cm}}The proposed model is better at detecting
activity in simulated data than standard models, without inflating
the false positive rate. When applied to real fMRI data, our GP model
in several cases finds brain activity where previously proposed LTI
models does not.\vspace{0.1cm}

\textbf{\hspace{-0.5cm}}\textbf{\noun{Conclusions}}

\textbf{\hspace{-0.5cm}}We have proposed a new non-linear model for
the hemodynamics in task fMRI, that is able to detect active voxels,
and gives the opportunity to ask new kinds of questions related to
hemodynamics.
\end{abstract}

\end{frontmatter}{}

\let\originalleft\left 
\let\originalright\right 
\renewcommand{\left}{\mathopen{}\mathclose\bgroup\originalleft} 
\renewcommand{\right}{\aftergroup\egroup\originalright}

\section{Introduction}

\subsection{Background}

Task based fMRI data are typically analyzed using voxelwise general
linear models (GLM), to detect voxels or regions where the blood oxygenation
level dependent (BOLD) contrast is correlated with the experimental
stimuli paradigm \citep{friston1994analysis,lindquist2008statistical}.
BOLD is an indirect measure of neural activation which depends on
the hemodynamic response (HR). Understanding the HR is therefore critical
in order to correctly infer the brain activity \citep{handwerker2004,lindquist2007validity,lindquist2009modeling}.
The neurovascular coupling between the neural response triggered by
a stimulus and the observed BOLD response in fMRI is not fully understood
\citep{logothetis2002,logothetis2003}, and the HR has been shown
to vary across voxels, brain regions and subjects \citep{handwerker2004,handwerker2012}.
It is common practice in fMRI to model the HR as a linear time invariant
system (LTI) \citep{boynton2012}. Standard GLMs make very strong
assumptions about the HR, and since it is unlikely that these models
are correct for all voxels and subjects, the inference for the brain
activity parameters will be biased \citep{lindquist2007validity,handwerker2004}.

\subsection{Joint Detection Estimation framework and Gaussian Processes}

The so called joint detection estimation (JDE) framework for the GLM
estimates the brain activity jointly with the HR. The JDE approach
uses a zero mean Gaussian process prior on the FIR filter coefficients,
which represent the HR in a LTI context. The filter is often called
the hemodynamic response function (HRF) \citep{goutte2000modeling,ciuciu2003,marrelec2003,casanova2008}.
In the FIR approach the HRF is convolved with the stimulus paradigm
to produce the predicted BOLD response, which is the idealized BOLD
response given neural activation. A problem with such voxelwise approaches
is that the filter is unidentified if the specific voxel is inactive,
since the filter has a zero mean prior on the HRF. There is also a
risk of overfitting, since a separate HR is estimated in each voxel.

Another approach is to use a bilinear model where both the design
matrix and the regression coefficients are estimated jointly. Many
models based on the JDE framework use parcellation, see for example
\citep{makni2008,vincent2010}. Some parameters are constant within
each parcel, while other parameters are voxel specific. Parcellation
can be done a priori and considered constant \citep{makni2008,vincent2010},
or estimated as a part of the model \citep{chaari2012hemodynamic,chaari:hal-01255465,albughdadi2017}.
By letting the HRF to be constant over a parcel reduces the amount
of overfitting that can happen compared to voxelwise approaches. The
parcellation approach in the JDE is a trade-off between gaining signal
to facilitate the estimation of the HRF, but still let the HRF vary
across the brain. For the approaches that use a fixed and a priori
known parcellation, it is assumed that there is \textit{one} HRF per
parcel, regardless of the number of stimuli in the experiment. The
idea is that a functionally similar region has the same hemodynamic
behavior.

\subsection{Non-linearity of Predicted BOLD}

There is evidence that contradicts the LTI system hypothesis for the
HR, see for example \citep{huettel2004functional} for a discussion.
This has motivated the development of more physiologically realistic
models that do not assume an LTI system, and model the predicted BOLD
directly \citep{buxton1998dynamics,friston1998,friston2000,buxton2004,deneux2006eeg,stephan2007comparing,lundengaard2016mechanistic}.
Estimation of such nonlinear models typically suffers from instabilities
and non-identifiability issues, and are more computationally expensive.
Non-linear extensions of JDE models that focus on the non-linear habituation
effect of repeated stimuli \citep{ciuciu2009modelling} are more efficient,
but accounts only for a limited class of non-linearities.

\subsection{The proposed approach}

In this work, we propose a new model that places a Gaussian Process
(GP) prior \citep{rasmussen2006} directly on the predicted BOLD response.
This is in contrast to earlier work which use a Gaussian process prior
on the HRF, and then convolve the prior HRF with the paradigm to obtain
the predicted BOLD response. Our approach is therefore not restricted
to LTI systems, which means that non-stationary and non-linear properties
of the BOLD response can be handled, if supported by the data. Non-stationarity
of the BOLD response can for example arise from refractory and adaptation
effects \citep{huettel2004functional}, or from a participant's failure
to perform a task in the MR scanner. Our approach can also implicitly
account for the so called stimulus-as-fixed-effect fallacy \citep{westfall2016fixing}.

A GP prior on the predicted BOLD makes the model very flexible, which
can lead to overfitting. Our model therefore incorporates several
features to avoid overfitting. First, we use a parcellation approach
similar to \citep{makni2008,vincent2010}, where the predicted BOLD
is restricted to be the same for all voxels in a given parcel, but
the activation and other parameters (for example time trends) are
voxel-specific. The effect is that the predicted BOLD in a parcel
is accurately estimated from data in many voxels. Second, in contrast
to the JDE literature, the mean of our GP prior is non-zero and equal
to the predicted BOLD from a physiologically motivated model of the
hemodynamics, for example the Balloon model proposed by \citep{buxton1998dynamics,buxton2004}.
This allows the GP posterior to fall back on the baseline physiological
model whenever the data are weak or support the baseline model, while
still being able to override the prior mean when the data are incompatible
with the baseline model. Third, using a well founded prior mean makes
it possible to use relatively tight priors on the GP.

The focus in this work is to detect voxel activity without being forced
to have restrictive assumptions about the dynamical system that drives
the hemodynamics of the BOLD response, but still be able to \textit{quantify}
the uncertainty of the response from that system. Section \ref{subsec:LTI-projected-posterior}
also proposes a projection of the GP posterior to a given LTI system
from which inferences on the usual HRF features such as time-to-peak,
time-to-undershoot etc can be obtained. Moreover, the residuals of
this projection provides information on where in time the LTI assumption
is likely to be violated, if at all.

The rest of the paper is organized as follows. Section 2 describes
the model and the inference procedure. Results from simulations and
real data are given in Section 3 and 4, respectively. The paper ends
with a discussion in Section 5 and conclusions in Section 6.

\section{Model and Bayesian inference}

\subsection{Notation}

Vectors and matrices are denoted with bold lower and upper case letters,
respectively. Vectors are assumed to be column vectors. The symbol
$^{\top}$ denotes transpose, $\mathbb{I}_{a}$ denotes the identity
matrix of size $a\times a$, $vec\left(\cdot\right)$ is the vectorization
operator, $\otimes$ is the Kronecker product and $diag\left(\boldsymbol{x}\right)$
means a diagonal matrix with vector $\boldsymbol{x}$ as the main
diagonal. $N\left(\boldsymbol{\mu},\boldsymbol{\Sigma}\right)$ and
$MN\left(\boldsymbol{\mu},\boldsymbol{\Sigma},\boldsymbol{\Omega}\right)$
denote multivariate normal and matrix normal (see Appendix A) distributions,
respectively. $InvGamma\left(a,b\right)$ denotes the inverse gamma
distribution. The following different indices are used:
\begin{itemize}
\item $j$: voxels, $j\in\{1,...,J\}$, within a parcel
\item $m$: stimulus type, $m\in\{1,\ldots,M\}$
\item $p$: number of nuisance variables, $p\in\{1,\ldots,P\}$
\item $k$: number of parameters in the $AR(k)$ process, $k\in\{1,...,K\}$.
\item $t$: time, $t\in\mathcal{T}_{\star}=\{-K+1,-K+2,\ldots,0,1,...,T\}.$
\end{itemize}

\subsection{Multivariate GLM for joint detection and estimation}

The fMRI signal will be modeled in the following way: hemodynamic
responses are the same for all voxels in a parcel, while task related
activations and parameters for the noise process are allowed to vary
between voxels in a parcel. The time series contain temporal autocorrelation
modeled by an autoregressive (AR) process of order $K$. We make the
usual simplifying assumption in time series analysis that the first
$K$ values of the process are known; let $\mathcal{T}_{0}=\{-K+1,-K+2,\ldots,0\}$
denote this initial set of time points. Further, define $\mathcal{T}=\{1,\ldots,T\}$
to be the subsequent time points and $\mathcal{T}_{\star}=\{\mathcal{T}_{0},\mathcal{T}\}$
to be the set of all $T_{\star}=T+K$ time points.

We use a multivariate regression model for all observed BOLD signals
during time $\mathcal{T}$ in one parcel, i.e.

\begin{equation}
\mathbf{Y}=H\left(\mathbf{F}\right)\mathbf{B}+\mathbf{Z}\boldsymbol{\Gamma}+\mathbf{U},\label{eq:full multivar model}
\end{equation}
where $\mathbf{Y}=\left(\begin{array}{ccc}
\mathbf{y}_{1} & \cdots & \mathbf{y}_{J}\end{array}\right)$ is a $T\times J$ matrix with the observed BOLD signal. The matrix
\textbf{$\mathbf{Z}$}, of size $T\times P$, contains nuisance covariates,
such as time trends and head motion covariates. $\mathbf{U}=\left(\begin{array}{ccc}
\mathbf{u}^{(1)} & \cdots & \mathbf{u}^{(J)}\end{array}\right)$ contains the model errors. The matrix $\mathbf{F}$ and its transformation
$H(\mathbf{F})$, which are explained in the next section, are both
of size $T\times M$. The idea is to let $H\left(\mathbf{F}\right)$
model the dynamics in the predicted BOLD response, while $\mathbf{B}$
models the overall response magnitude in each voxel, of size $M\times J$.
The separation of the hemodynamics from the activations gives a straight
forward measure of voxel activation, that can be used to construct
posterior probability maps (PPMs) or $t$-maps.

We assume that the noise in each voxel follows an $AR(K)$ process,
i.e. for the $j$th voxel
\[
\mathbf{u}^{(j)}=\rho_{1}\mathbf{u}_{-1}^{(j)}+\rho_{2}\mathbf{u}_{-2}^{(j)}+\ldots+\rho_{k}\mathbf{u}_{-k}^{(j)}+\boldsymbol{\epsilon}^{(j)},\:j=1,...J,
\]
where the negative indices denote time lags. We assume that the AR
parameters are the same for all voxels in a parcel, but different
across parcels. The error terms $\boldsymbol{\epsilon}^{(j)}$ are
assumed to be independent across voxels and $\boldsymbol{\epsilon}^{(j)}\sim N\left(\mathbf{0},\sigma_{j}^{2}\mathbb{I}_{T}\right)$.
The distribution of $\mathbf{U}$ can be expressed as

\[
vec(\mathbf{U})\sim N\left(\mathbf{0},\boldsymbol{\Omega}\otimes\mathbf{M}_{\boldsymbol{\rho}}\right),
\]
where $\boldsymbol{\Omega}=diag\left(\boldsymbol{\sigma}^{2}\right)$
and $\boldsymbol{\sigma}^{2}=(\sigma_{1}^{2},...,\sigma_{J}^{2})$.
The matrix $\mathbf{M}_{\boldsymbol{\rho}}$ can be obtained by solving
a system of Yule-Walker equations or using the methods of \citep{van1994covariance},
but it is not needed explicitly for sampling from the posterior of
the model in (\ref{eq:full multivar model}). Spatial noise dependencies
can also be incorporated by letting $\boldsymbol{\Omega}$ be a matrix
that reflects the spatial distance between voxels.

\subsection{A physiological Gaussian process prior for predicted BOLD\label{subsec:A-physiological-Gaussian}}

The predicted BOLD is modeled with a Gaussian process (GP) prior \citep{rasmussen2006}
according to
\[
f(t)\sim GP\left(m\left(t,\boldsymbol{\xi}\right),\mathtt{k}\left(t,t',\boldsymbol{\theta}\right)\right),
\]
where $m\left(t,\boldsymbol{\xi}\right)$ is the mean function with
hyperparameter $\boldsymbol{\xi}$, and $\mathtt{k}\left(t,t',\boldsymbol{\theta}\right)$
is the kernel (covariance function) with hyperparameters $\boldsymbol{\theta}$.
The stimuli paradigm affects the process through the prior mean function.
A sampled value of the GP is denoted $f_{t}$ and 
\[
\mathbf{f}=\left(\begin{array}{cccc}
f_{-k+1} & f_{-k+2} & \cdots & f_{T}\end{array}\right)^{\top}.
\]
This prior gives a general framework for modeling the hemodynamic
response with a variety of physiological models or constraints. The
mean function, defined by the parameters $\boldsymbol{\xi}$, can
come from some arbitrary model that can generate the predicted BOLD.
These models can be linear (e.g. the HRF used in the SPM software)
or non-linear (e.g. the Balloon model). The kernel controls both the
degree of smoothness of $\mathbf{f}$ and the deviation from the mean
function. If several stimuli are considered, the total predicted BOLD
is considered to be a linear combination of different GPs, which will
be denoted $\mathbf{f}_{m}$, and the $(T+K)\times M$ matrix $\mathbf{F}=\left(\begin{array}{ccc}
\mathbf{f}_{1} & \cdots & \mathbf{f}_{M}\end{array}\right)$ gather all the sampled realizations of the different GPs over all
time periods $\mathcal{T}_{\star}$. The parameters $\boldsymbol{\xi}_{m}$
and $\boldsymbol{\theta}_{m}$ denote the stimulus specific GP hyper-parameters.

We use a Matérn kernel with $\frac{5}{2}$ degrees of freedom
\[
\begin{array}{c}
\mathtt{k}_{\nu=5/2,m}\left(r\right)=\omega_{m}^{2}\left(1+\frac{\sqrt{5}r}{l_{m}}+\frac{5r^{2}}{l_{m}^{2}}\right)exp\left(-\frac{\sqrt{5}r}{l_{m}}\right)\end{array},
\]
where $r=\left\Vert t_{1}-t_{2}\right\Vert $ is the Euclidean distance
between two covariate observations. We use time as kernel covariate,
but it is possible to use other covariates. The lengthscale parameter
$l$ controls the degree of smoothness of the predicted BOLD and $\omega_{m}^{2}$
the prior variability around the mean $m\left(t,\boldsymbol{\xi}\right)$.
The traditional model with fixed predicted BOLD is obtained by letting
$\omega_{m}^{2}\rightarrow0$. Note that although this prior is stationary,
the posterior need not be, see Figure \ref{fig:sub19_pbold_157} for
an example. Let $\boldsymbol{\theta}_{m}=\left(\begin{array}{cc}
l_{m} & \omega_{m}^{2}\end{array}\right)$, $\boldsymbol{\theta}=(\boldsymbol{\theta}_{1},...,\boldsymbol{\theta}_{M})$
and denote the covariance matrix for all data points from $\mathtt{k}\left(t,t',\boldsymbol{\theta}_{m}\right))$
by $\mathbf{\mathtt{K}}\left(\mathcal{T}_{\star},\mathcal{T}_{\star}\right)_{m}.$
The hyperparameters for the kernel are for simplicity assumed to be
fixed and known, but can in principle be learned in separate updating
steps; see the discussion in Section \ref{subsec:Future-work}.

\subsection{Identifying restrictions on the hemodynamics}

Modeling the voxel activation as $\mathbf{F}\mathbf{B}$ has the drawback
that the parameters enter the likelihood as a product, and are thereby
not individually identified since $\mathbf{F}\mathbf{B}=\mathbf{F}\boldsymbol{S}^{-1}\boldsymbol{S}\mathbf{B}$,
for any invertible matrix $\boldsymbol{S}$ of size $M\times M$.
To overcome this problem, we propose an identifying nonlinear transformation
for the matrix $\mathbf{F}$. The transformation must fixate the scale
of $\mathbf{F}$, and prevent sign flipping as well as linear combinations
and permutations. Similar to \citep{pedregosa2015data}, we propose
the transformation:

\begin{equation}
h\left(\mathbf{f}_{m}\right)=\mathbf{f}_{m}/\left(||\mathbf{f}_{m}||_{\infty}\mathrm{sign}\left(\mathbf{f}_{m}^{\top}m\left(t,\boldsymbol{\xi}\right)\right)\right).\label{eq:first trans}
\end{equation}

This transformation is still sensitive to pure permutations, so in
order to identify the column order we introduce a permutation function
$\Psi\left(\cdot\right)$. Let $\mathbf{F}_{0}$ be prior mean for
$\mathbf{F}$ and $\mathbf{F}_{post}$ be a sample from the posterior
for $\mathbf{F}$. The function $\Psi\left(\mathbf{F}_{post}\right)$
permutes the column in $\mathbf{F}_{post}$ in such a way that

\begin{equation}
\sum_{m=1}^{M}\mathrm{corr}\left(\mathbf{f}_{post,m}\mathbf{f}_{0,m}\right),\label{eq:corr-perm_func}
\end{equation}
is maximized with respect to the column order of $\mathbf{F}_{post}$.
The interpretation of the function $\Psi\left(\cdot\right)$ is the
following: If there is an ambiguity which column position the vector
$\mathbf{f}_{m}$ shall have then it shall be placed in that column
in which it has the highest correlation with the column specific prior
mean. The final transformation is then given by
\begin{equation}
H\left(\mathbf{F}\right)=\Psi\left(\left[h\left(\boldsymbol{\mathbf{f}}_{1}\right),\ldots,h\left(\boldsymbol{\mathbf{f}}_{M}\right)\right]\right).\label{eq: trans final}
\end{equation}
$H\left(\mathbf{F}\right)$ has its support on a relative scale and
is bounded between -1 and 1. The use of the function $H\left(\mathbf{F}\right)$
is possible due to the use of an informative prior mean for $\mathbf{F}$.

\subsection{Posterior computations}

The likelihood and the prior on all parameters are described in Appendix
B. The joint posterior of all model parameters in (\ref{eq:full multivar model})
is given by

\begin{equation}
\begin{array}{c}
p\left(\mathbf{F},\mathbf{B},\boldsymbol{\Gamma},\boldsymbol{\sigma}^{2},\boldsymbol{\rho}|\mathbf{Y},\mathbf{Z}\right)\propto L\left(\mathbf{Y}|\mathbf{F},\boldsymbol{\Gamma},\mathbf{B},\boldsymbol{\sigma}^{2},\boldsymbol{\rho},\mathbf{Z}\right)\\
\times p\left(\mathbf{F}\right)p\left(\mathbf{B}\vert\boldsymbol{\sigma}^{2}\right)p\left(\boldsymbol{\Gamma}\vert\boldsymbol{\sigma}^{2}\right)p\left(\boldsymbol{\sigma}^{2}\right)p\left(\boldsymbol{\rho}\right).
\end{array}\label{eq:joint posterior}
\end{equation}
Equation (\ref{eq:joint posterior}) is intractable and cannot be
sampled using standard distributions, and we therefore resort to Gibbs
sampling. The posterior is sampled in four steps, which are described
in Algorithm \ref{alg:Mini Gibbs} and detailed formulas are given
in the Appendix C.

\begin{algorithm}
\begin{enumerate}
\item $\boldsymbol{\rho}$ is sampled from a multivariate Gaussian distribution
in Equation (\ref{eq:rho_full_cond_post}). $\boldsymbol{\rho}$ is
then used to pre-whiten the data before sampling of the other parameters.
\item $\boldsymbol{\sigma}^{2}$ is sampled from an Inverse-gamma distribution
in Equation (\ref{eq:posterior sigma-1}).
\item $vec\left(\boldsymbol{B}\right)$ and $vec\left(\boldsymbol{\Gamma}\right)$
are sampled from a multivariate normal distribution in Equation (\ref{eq: posterior B conditional omega-1}).
\item \textbf{$\mathbf{F}$ }is sampled with Elliptical slice sampling using
the likelihood in (\ref{eq: GP likelihood}).
\end{enumerate}
\caption{Schematic description of the Gibbs sampler for the posterior in Equation
(\ref{eq:joint posterior}).\label{alg:Mini Gibbs}}
\end{algorithm}

\subsection{LTI projected posterior\label{subsec:LTI-projected-posterior}}

Some hemodynamic features can be investigated manually by inspecting
the posterior distribution of $\mathbf{F}$, see for example the results
in Figure \ref{fig:sub19_pbold_157}. We will here propose a decomposition
of the posterior for $\mathbf{F}$ into a linear LTI part and a non-linear
part, by projecting the GP posterior to the closest LTI model. Let
$\tilde{\mathbf{f}}_{m}^{(q)}$ be the $q$:th posterior draw of the
$m$:th stimulus and compute the projection
\[
\mathbf{h}_{m}^{(q)}=\left(\mathbf{X}_{FIR}^{\top}\mathbf{X}_{FIR}\right)^{-1}\mathbf{X}_{FIR}^{\top}\mathbf{f}_{m}^{(q)},
\]
where $\mathbf{X}_{FIR}$ is the standard FIR design matrix for one
stimulus and $\boldsymbol{h}_{m}$ contains the $K$ FIR projected
filter coefficients. The $\mathbf{h}_{m}^{(q)}$ are samples from
the FIR projected posterior from which we can directly compute the
posterior for HRF features such as time-to-peak, time-to-undershoot
etc. Moreover, the residuals
\[
\epsilon_{f,m}^{(q)}=\mathbf{f}_{m}^{(q)}-\mathbf{X}_{FIR}\mathbf{h}_{m}^{(q)},
\]
are time varying deviation from the LTI system. The posterior of the
$\epsilon_{f,m}^{(q)}$ therefore provide information on which time
periods show the largest deviations from the LTI model.

\subsection{Implementation}

The proposed estimation of the model in (\ref{eq:full multivar model})
is implemented in the R programming language. Since each parcel is
independent, the computation can be parallelized across parcels, which
is done with the foreach package \citep{analytics2015foreach}. Functions
from the R package neuRosim \citep{welvaert2011} are used to simulate
data for the simulation study.

\section{Simulations\label{sec:Simulations}}

Most task fMRI experiments model the hemodynamics using one or a few
basis functions that are fixed across voxels and subjects. These assumptions
are likely to be wrong \citep{handwerker2004,lindquist2007validity,lindquist2009modeling}
and lead to some degree of model misspecification. We here perform
a simulation study to investigate the proposed model's ability to
detect activity in different scenarios and assess the sensitivity
of model misspecification.

Simulated data were generated from the generative model in equation
\ref{eq:full multivar model}. In each simulation a single parcel
with 100 voxels was used, and 20 of the voxels were active. The following
settings were used for each simulation
\begin{itemize}
\item A single stimulus was modeled as a block paradigm. The number of time
points was set to 150 and the sampling rate (TR) was set to 1 second.
\item An AR(3) process was used for the noise process, with autoregressive
parameters: $\boldsymbol{\rho}^{T}=(0.4,0.1,0.05).$
\item The contrast to noise ratio (CNR), $b_{m,j}/\sigma$, where $b_{m,j}=1$
for active voxels and zero otherwise, was set to 5 and 7.
\item Constant, linear, quadratic and cubic trends were added to each time
series. The coefficients were generated randomly for each voxel and
simulation, to reflect realistic trends in fMRI data.
\end{itemize}
We fit two models to each dataset. In the first model, the predicted
BOLD is fixated to the double gamma HRF convolved with the paradigm.
In the second model, the predicted BOLD is estimated from the data
using our physiological GP prior where the double gamma HRF convolved
with the paradigm is now the prior mean. The prior mean for $\mathbf{F}$
was created in such a way that is was either correct, with correlation
1 with the true process, or erroneous with a correlation of 0.615
with the true process. This allows us to assess the sensitivity of
model misspecification.

The two levels of CNR and the two settings for predicted BOLD (correct
or not correct) give us a total of four different simulation setups.
We generated 32 datasets for each of the four setups. The same priors
were used for all simulations and for the real data and were specified
as given in Appendix B. We use two different length scales for the
GP prior, $l=2$ or $l=4$, and a GP standard deviation of $\omega=\sqrt{0.1}\approx0.316$.

The posterior was sampled 4000 times and 1000 samples were discarded
as burn in. The remaining samples were thinned out by a factor of
3, leaving 1000 posterior draws for inference.

There are several ways that voxels can be declared active in an Bayesian
model. One way is to construct posterior probabilities of the type
$p\left(b_{m,j}>c\vert\mathbf{y}\right)>a$, where $c$ is a user
defined effect size and $a$ is a probability threshold. These probabilities
can be used to construct PPMs over the brain. However, there is no
consensus regarding how to threshold PPMs, and PPMs can be numerically
unstable and imprecisely estimated. Instead, we use Bayesian ``$t$-ratios'',
defined as
\[
t=\frac{\mathrm{E}(b\vert\mathbf{y})-c}{\sqrt{\mathrm{Var}(b\vert\mathbf{y})}},
\]
which can be easily computed from the posterior samples in each voxel.
These ratios have a higher resolution for a fixed amount of posterior
samples compared to PPMs. In the simulations we set $c=0$ and the
test $t>a$ was used, where $a$ is a quantile from a $t$-distribution.
Note that the activation is voxel independent given $\mathbf{F}$
and $\boldsymbol{\rho}$, due to the simulation design. These frequentist
calculations are of course not directly transferable to a Bayesian
setting, but has the advantage of giving familiar thresholds for fMRI.

\begin{figure}
\includegraphics[scale=0.4]{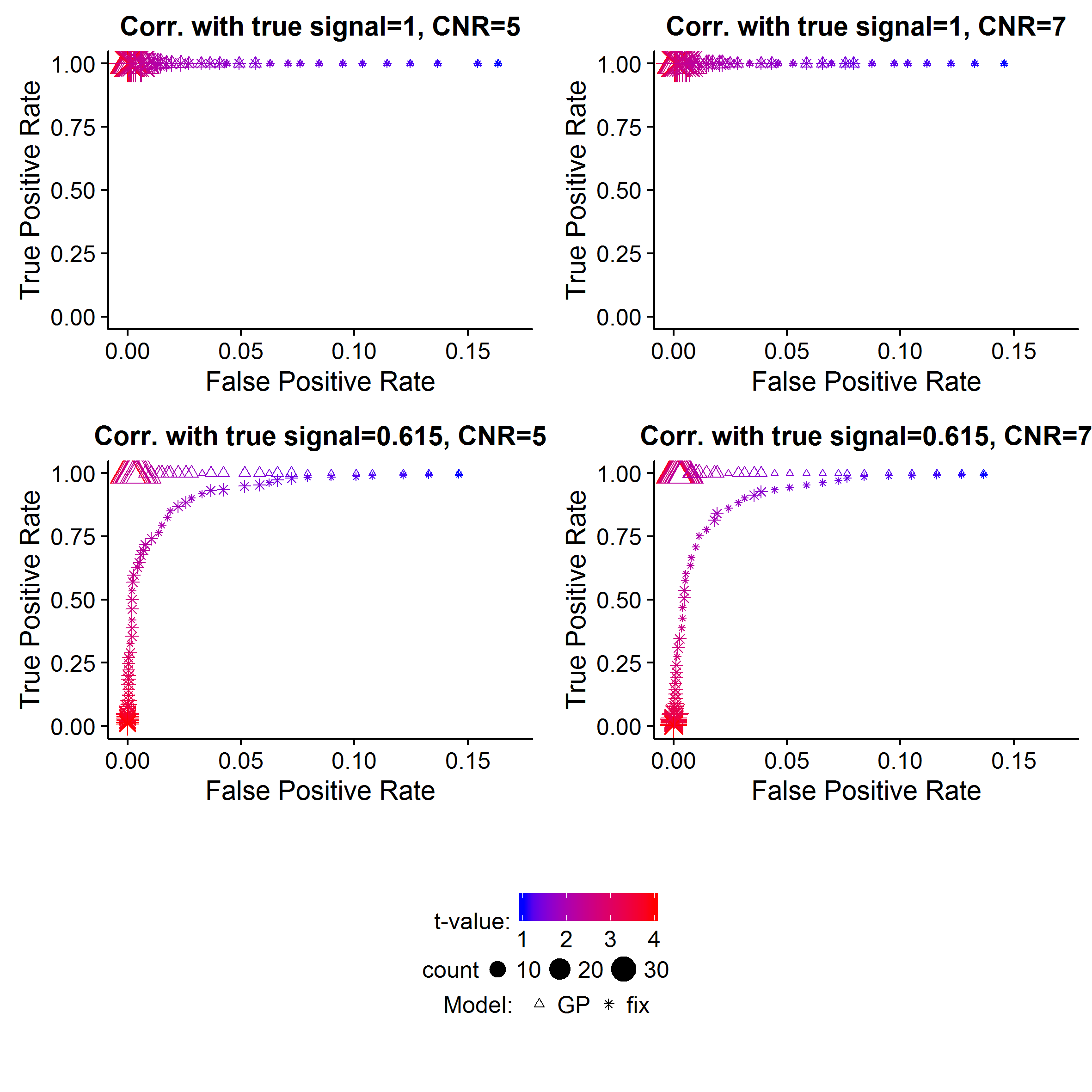}\caption{ROC curves for the simulation study. Lengthscale hyperparameter for
GP: $l=4$. Each value of is the average of over 32 simulations. Thresholds
for $t$-values: 60 equidistant values between 1 and 4. \label{fig:ROC_GP_l=00003D4}}
\end{figure}

We compute the test $t>a$ for a range of 60 equidistant values for
$a$ between 1 and 4, and compute the true positive rates and false
positive rates over the whole parcel and then averaged over the simulated
datasets. Figure \ref{fig:ROC_GP_l=00003D4} shows the resulting ROC
curves for the lengthscales $l=4$. The results of $l=2$ is are very
similar to those of $l=4$ and are not shown. The subplots show the
results for all the combinations of simulation designs. The count
variable in the plots indicate the number of $a$:s that approximately
have the same value for the false positive rate. The top row shows
that our model with estimated predicted BOLD does as well as the fixed
predicted BOLD when the correct predicted BOLD is used for the fixed
model. The bottom row shows that estimating the predicted BOLD substantially
outperforms the misspecified fixed model, even when the same misspecification
is used as prior mean for the GP. As expected, for very low thresholds
the model with estimated predicted BOLD starts to classify non-active
voxels as active.

\begin{figure}
\includegraphics[scale=0.48]{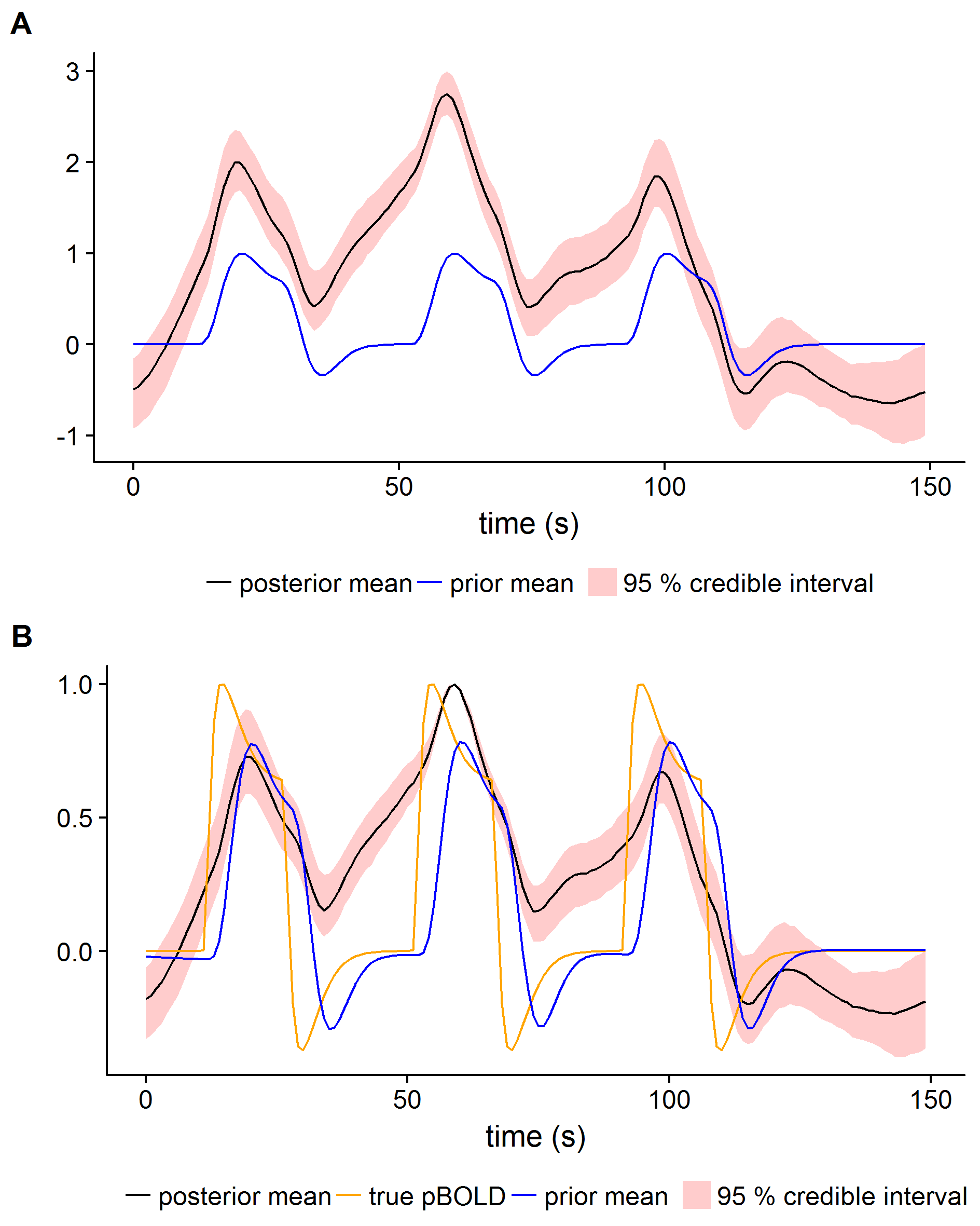}\caption{Posterior for predicted BOLD for a simulated parcel. CNR=5, correlation
with true signal is 0.615. Lengthscale hyperparameter for GP: $l=4$.
(A) is the Gaussian process $\mathbf{F}$ in \ref{eq:full multivar model}
and (B) is the transformed Gaussian process $H\left(\mathbf{F}\right)$,
see Equation (\ref{eq:first trans}) and (\ref{eq: trans final}).\label{fig: simulated data Predicted-BOLD parcel 8}\label{fig:sim_pbold_l=00003D4}}
\end{figure}

\begin{figure}
\includegraphics[scale=0.48]{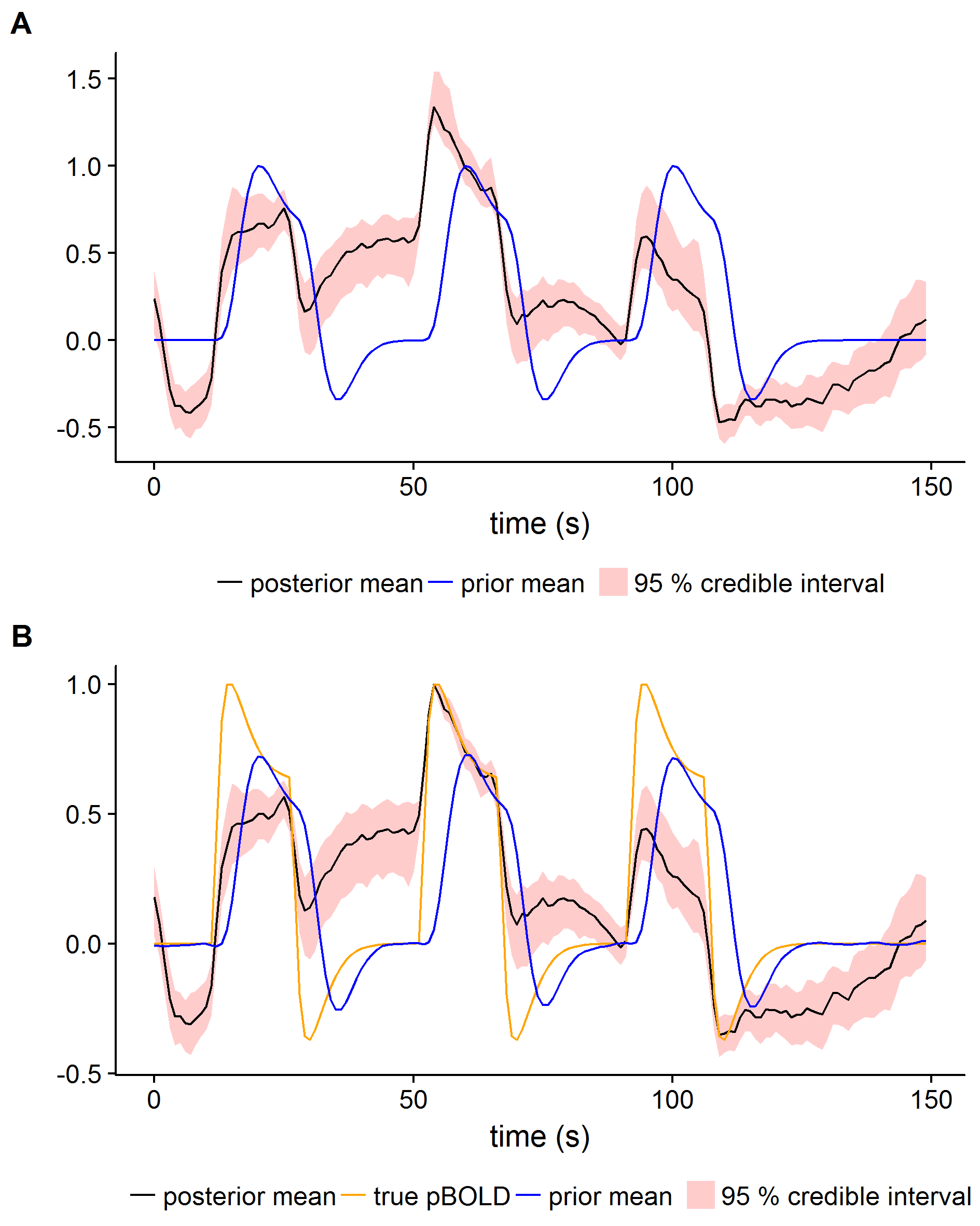}\caption{Posterior for predicted BOLD for a simulated parcel. CNR=5, correlation
with true signal is 0.615. Lengthscale hyperparameter for GP: $l=2$
. (A) is the Gaussian process $\mathbf{F}$ in \ref{eq:full multivar model}
and (B) is the transformed Gaussian process $H\left(\mathbf{F}\right)$,
see Equation (\ref{eq:first trans}) and (\ref{eq: trans final}).
\label{fig: simulated data Predicted-BOLD parcel 8-5}\label{fig:sim_pbold_l=00003D2}}
\end{figure}

Figure \ref{fig: simulated data Predicted-BOLD parcel 8} and \ref{fig: simulated data Predicted-BOLD parcel 8-5}
show the posterior of the predicted BOLD for selected simulations. 

Our application on real data in Section \ref{sec:Real-data} shows
that the flexibility of the proposed GP model makes it possible to
find more activity than with a LTI model for the hemodynamics. It
could of course be that the GP model overfits the data and finds activity
where there actually are none. To explore this we simulate data from
an additional scenario where there is no underlying activity, and
the models are estimated with a simple made-up block paradigm. Table
\ref{tab:NoActivity} reports the percentage of false positives using
the rule $|t|>2$ to declare a voxel as active. Bayesian $t$-ratios
do not have controlled false positive rates so the expected false
positive rate is not known, and we do not control for multiple testing;
but since we are interested in the comparison of false positive rates
\emph{relative} to the model with fixed HRF, this is not a concern
here. Table \ref{tab:NoActivity} shows that the false positive rates
of the GP and fixed HRF models are comparable when the length scale
is $l=4$; the shorter length scale seems to allow for too much flexibility
and therefore a somewhat inflated false positive rate. These results
agree with Figure \ref{fig:t-maps_real_lengthscale} for the real
data in Section \ref{sec:Real-data}, showing the importance of not
setting the length scale too small.

\begin{table}
\begin{centering}
\begin{tabular}{cccc}
\hline 
St.dev noise & Fixed HRF & GP $l=4$ & GP $l=2$\tabularnewline
\hline 
\multirow{1}{*}{0.33} & 3.3\% & 3.5\% & 6.7\%\tabularnewline
\multirow{1}{*}{0.20} & 3.3\% & 3.4\% & 6.6\%\tabularnewline
\multirow{1}{*}{0.14} & 3.3\% & 3.4\% & 6.3\%\tabularnewline
\hline 
\end{tabular}
\par\end{centering}
\caption{Percentage of false positives in the simulated data with no activity
using the rule $|t|>2$ to declare a voxel as active.\label{tab:NoActivity}}
\end{table}

\section{Real data\label{sec:Real-data}}

\subsection{Data}

To test our proposed approach on real data, we used open fMRI data
from brain tumor patients \citep{pernet,gorgolewski2013test}, as
the hemodynamic response function may be different close to a tumor.
A total of 22 patients (9 females) with different types of brain tumors
were scanned using both structural (T1, T2, DWI) and functional (BOLD
T2{*}) MRI sequences. For the functional scans several tasks were
performed: motor, verb generation and word repetition (resting state
data are also available). Data were acquired on a General Electric
1.5 Tesla scanner with an 8 channel phased-array head coil. The fMRI
data were acquired using a standard EPI sequence with a repetition
time of 5.0 seconds (due to sparse sampling for auditory tasks) or
2.5 seconds, and an echo time of 50 milliseconds. Each voxel has a
size of 4 x 4 x 4 mm$^{3}$, resulting in volumes with 64 x 64 x 30
voxels.

We here focus on the word repetition task. The task is to repeat a
given word (overt word repetition), in 6 blocks with 30 seconds of
activation and 30 seconds of rest. We can thereby expect activation
of the language areas of the brain, parts of the motor cortex that
correspond to the mouth and tongue (speech production) and the auditory
cortex (listening). Our presented results are for two randomly selected
subjects: 18716 and 19628. 

\subsection{Preprocessing}

The fMRI data were preprocessed using motion correction and 6 mm smoothing.
The brain parcellation was performed by registering the ADHD 200 parcel
atlas \citep{craddock,bellec2017neuro} to EPI space, by combining
linear T1-MNI and EPI-T1 transformations using FSL. A brain mask is
applied in order to remove voxels outside the brain. The number of
parcels can therefore differ across subjects as some parcels are removed
by the mask.

\begin{figure}
\centering{}\includegraphics[scale=0.33]{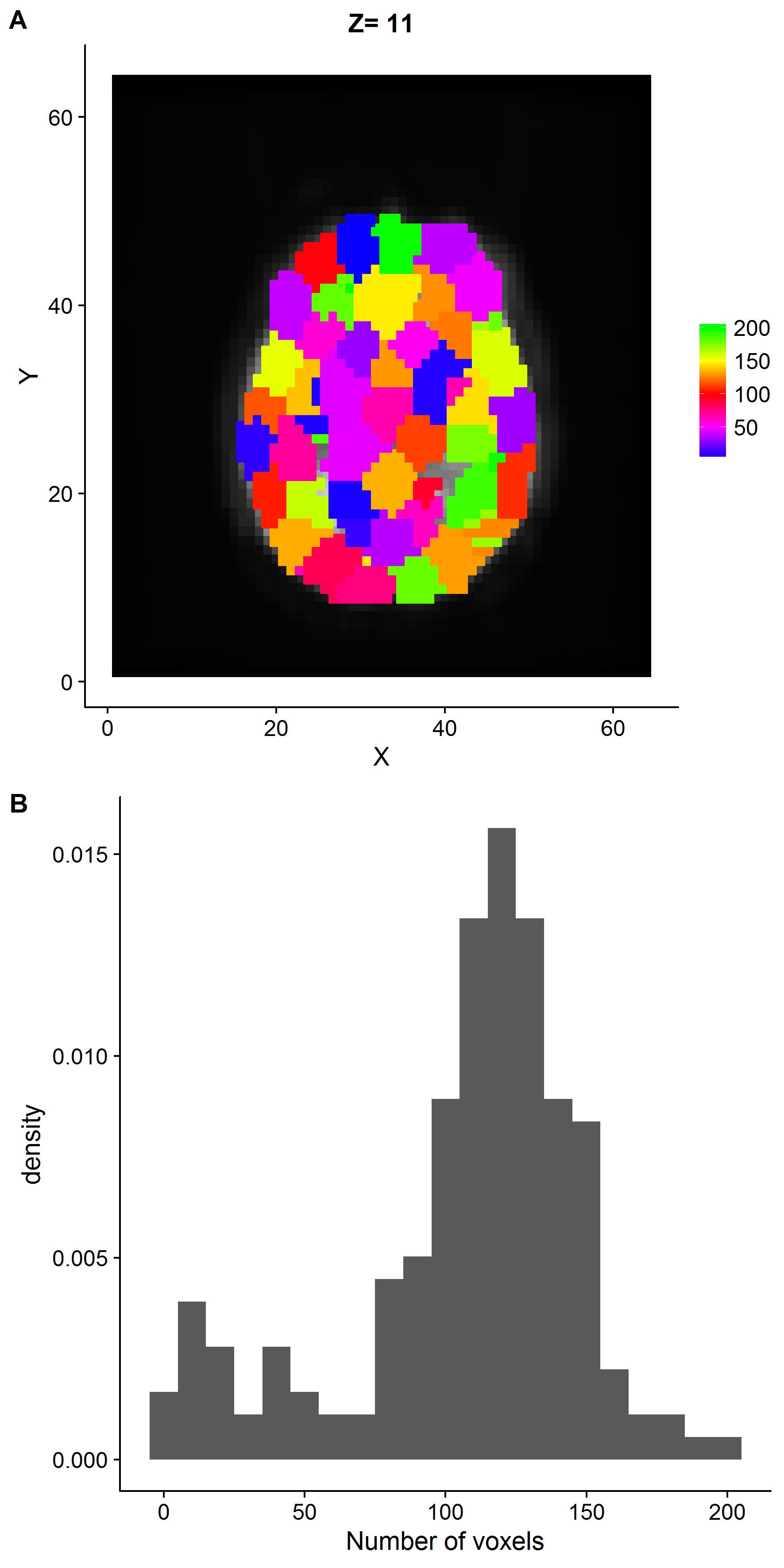}\caption{Descriptive statistics for subject 19628. (A) shows all parcels for
Z-slice 11. The color specifies parcel belonging. (B) shows the number
of voxels in all parcels. Note that our model falls back on the prior
mean if there are few active voxels in a parcel, meaning that the
model will not break down for small parcels. \label{fig:parcel stats}}
\end{figure}

In order to be able to compare effect sizes between voxels, the fMRI
data is scaled. We use the scaling 

\[
\frac{\boldsymbol{y}_{j}}{sd\left(\boldsymbol{y}_{j}\right)}\cdot\frac{100}{GM},
\]
where $\boldsymbol{y}_{j}$ is observed data for voxel $j$ , $sd\left(\cdot\right)$
is the standard deviation function and $GM$ is the global mean of
$\boldsymbol{y}_{j}/sd\left(\boldsymbol{y}_{j}\right)$ over all voxels.
This scaling ensures that overall the voxels have an average value
close to 100 and the same standard deviation. This makes it possible
to use effect sizes in terms of percent of the global mean signal,
as done in \citep{penny2005bayesian,siden2017fast}. There are 186
parcels and a total of 19,836 voxels for subject 18716. Corresponding
numbers for subject 19628 are 179 and 19,304, respectively. Some example
parcels and the distribution of parcel size for subject 19628 are
shown in Figure \ref{fig:parcel stats}. 

\subsection{Results}

Independent models were fitted to each parcel. An autoregressive model
of order 3 was used for the noise process. The same priors were used
for all parcels. The same prior hyperparameters were used as in the
simulation study except for $\mathbf{F}$, which used kernel hyperparameters
$l=4$ and $\omega=0.1$. The prior mean function for $\mathbf{F}$
was specified to the default HRF in SPM convolved with the paradigm,
and was scaled to have zero mean and unit variance.

Constant, linear, quadratic and cubic trends were included as nuisance
variables. See Appendix D for the details about the starting values
for the MCMC.

The proposed GP model is compared with three baseline models. In the
first model the predicted BOLD is fixated to the prior mean. The second
model also uses the temporal derivative of the prior mean, i.e. two
basis functions (which is the most common way to allow for a small
time shift of the paradigm). The third model uses a smooth FIR approach
\citep{goutte2000modeling,ciuciu2003,marrelec2003,makni2008,vincent2010}
to model the predicted BOLD (see Appendix E for details). The same
priors are used for all other parameters.

The posterior was sampled 9000 times, and 3000 samples were discarded
as burn in. The remaining samples were thinned out by a factor 6,
leaving a final sample of 1000 posterior draws for inference. The
parameters $\mathbf{B}$, $\mathbf{\boldsymbol{\Gamma}}$, $\boldsymbol{\sigma}^{2}$
and $\boldsymbol{\rho}$ showed a good mixing with low autocorrelation.
In some parcels, the elements in $\mathbf{F}$ had high autocorrelation,
but the first and second half of the draws gave similar posteriors.

\begin{figure*}
\centering{}\includegraphics[width=14.7273cm,height=9cm]{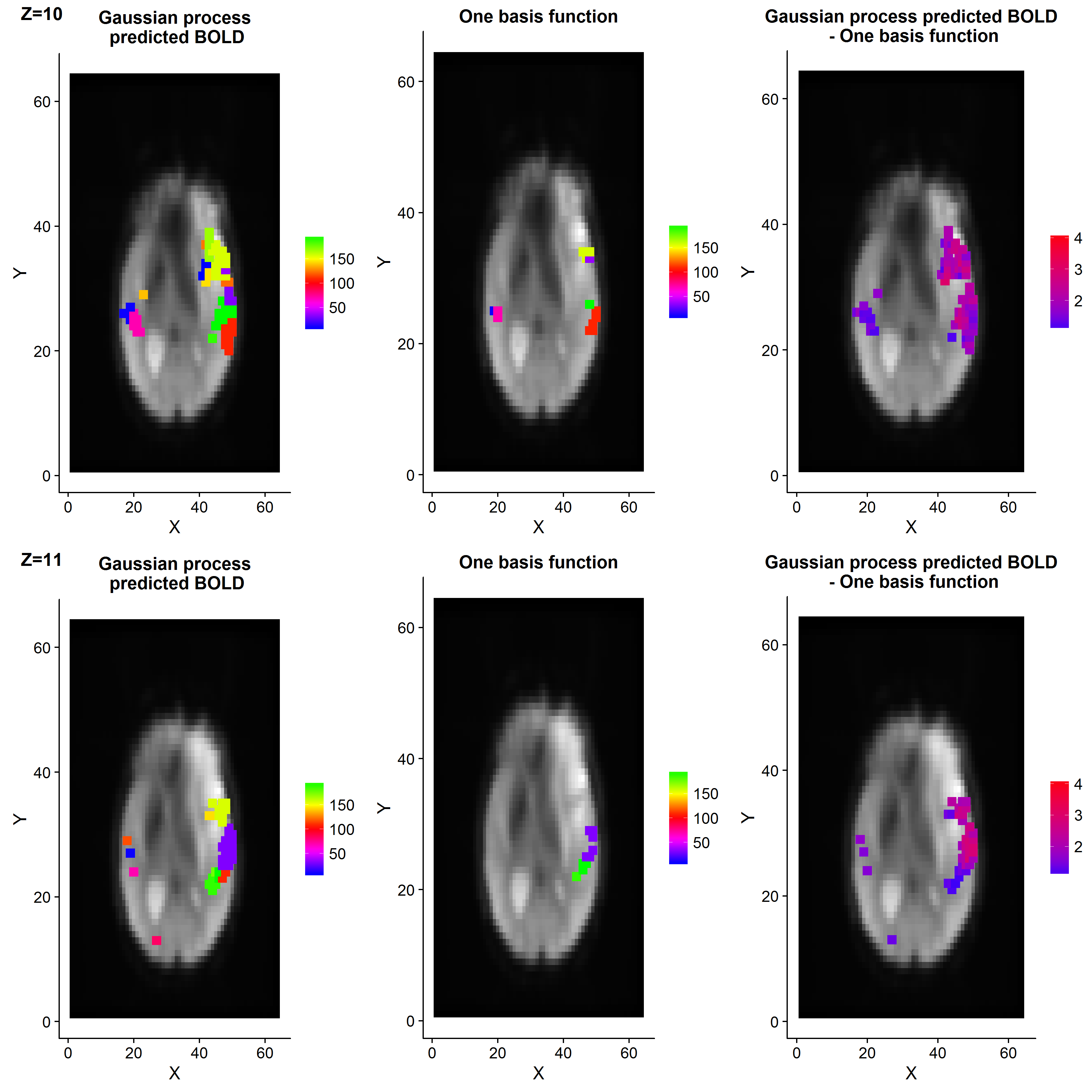}\caption{Example slices with Bayesian $t$-ratios for subject 19628. The activity
maps are thresholded at $t\ge4$ for a test that tests the effect
size $0.25$. The color specifies parcel belonging for active voxels.
The rightmost column shows the differences in t-ratios, thresholded
such that only values fulfilling $|t_{1}-t_{2}|>1$ are shown. Our
flexible model clearly detects more activity, compared to the fix
predicted BOLD model. Top row: the flexible model detects more brain
activity in Broca's language area, which for this subject is close
to the brain tumor. Bottom row: the flexible model finds brain activity
in the visual cortex and bilateral activation of the auditory cortex,
which the fix model struggles to detect. \label{fig:t-maps_real_one_basis}}
\end{figure*}

\begin{figure*}
\centering{}\includegraphics[width=14.7273cm,height=9cm]{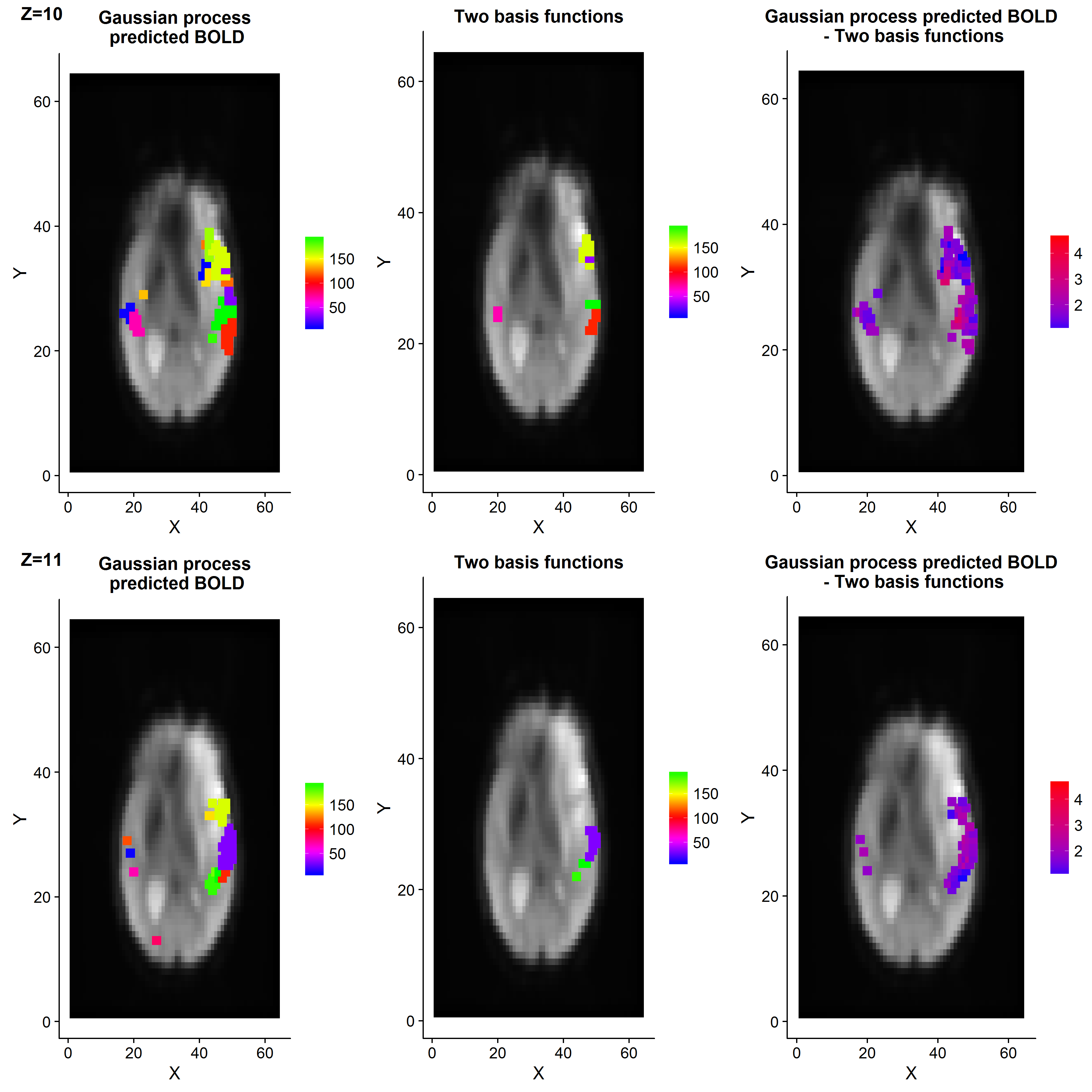}\caption{Example slices with Bayesian $t$-ratios for subject 19628. The activity
maps are thresholded at $t\ge4$ for a test that tests the effect
size $0.25$. t-ratios for the baseline model is created with $b$
for the first basis. The color specifies parcel belonging for active
voxels. The rightmost column shows the differences in t-ratios, thresholded
such that only values fulfilling $|t_{1}-t_{2}|>1$ are shown. The
differences between the flexible model and the fix model are very
similar to Figure \ref{fig:t-maps_real_one_basis}, where a single
basis function was used. \label{fig:t-maps_real_two basis}}
\end{figure*}

\begin{figure*}
\centering{}\includegraphics[width=14.7273cm,height=9cm]{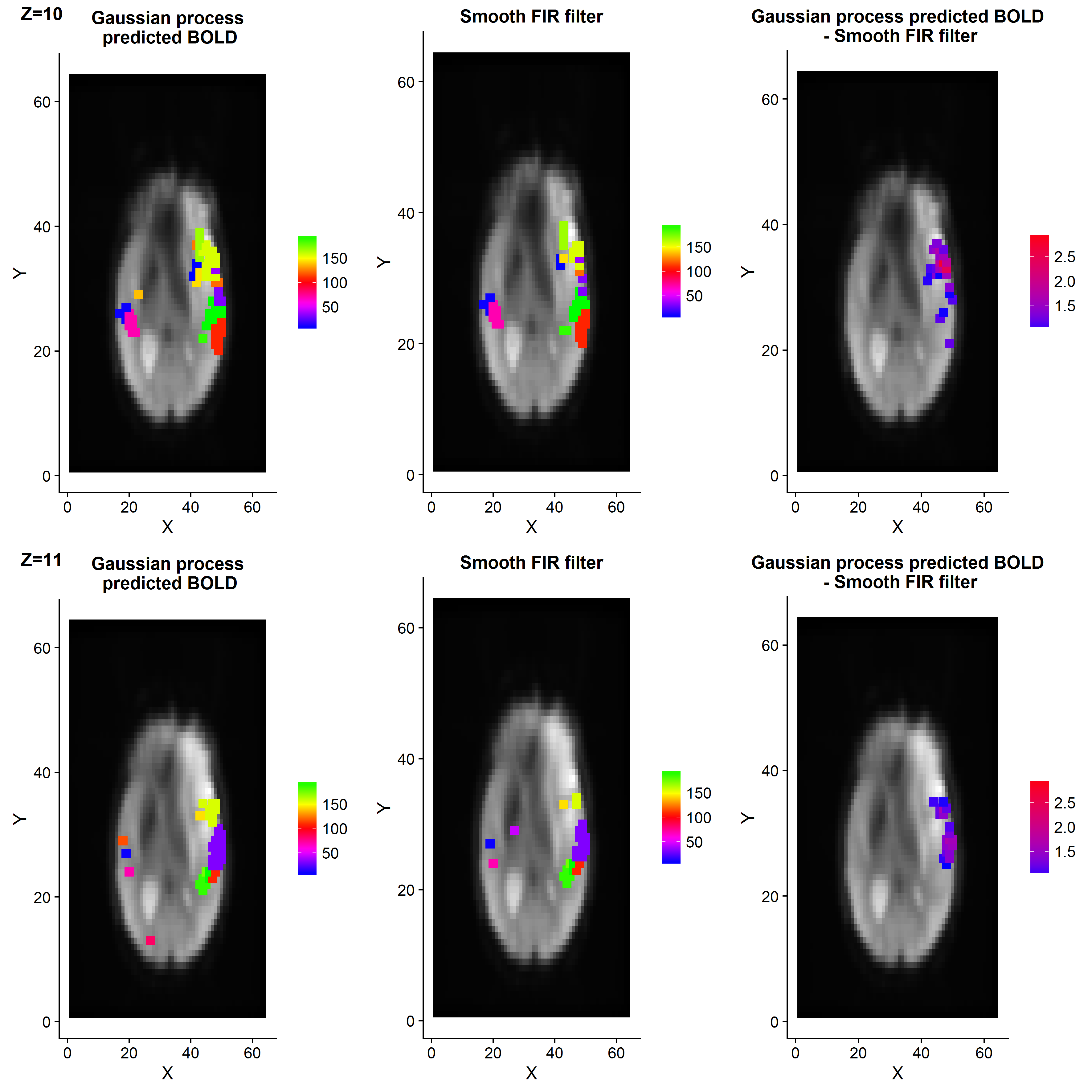}\caption{Example slices with Bayesian $t$-ratios for subject 19628. The activity
maps are thresholded at $t\ge4$ for a test that tests the effect
size $0.25$. The color specifies parcel belonging for active voxels.
The rightmost column shows the differences in t-ratios, thresholded
such that only values fulfilling $|t_{1}-t_{2}|>1$ are shown. \label{fig:t-maps_real_FIR_19}}
\end{figure*}

There can be sizeable differences in activation from using a GP prior
on the predicted BOLD, compared to using a fixed predicted BOLD. For
example, focusing first on subject 19628, Figures \ref{fig:t-maps_real_one_basis}
and \ref{fig:t-maps_real_two basis} show that the model that estimates
the predicted BOLD finds more activity compared to the two first baseline
models. For example, the flexible model detects more brain activity
in Broca's language area, which for this subject is close to the brain
tumor. According to Figure \ref{fig:t-maps_real_FIR_19}, the flexible
GP model also detects stronger brain activity compared to the smooth
FIR filter approach.

\begin{figure}
\includegraphics[scale=0.48]{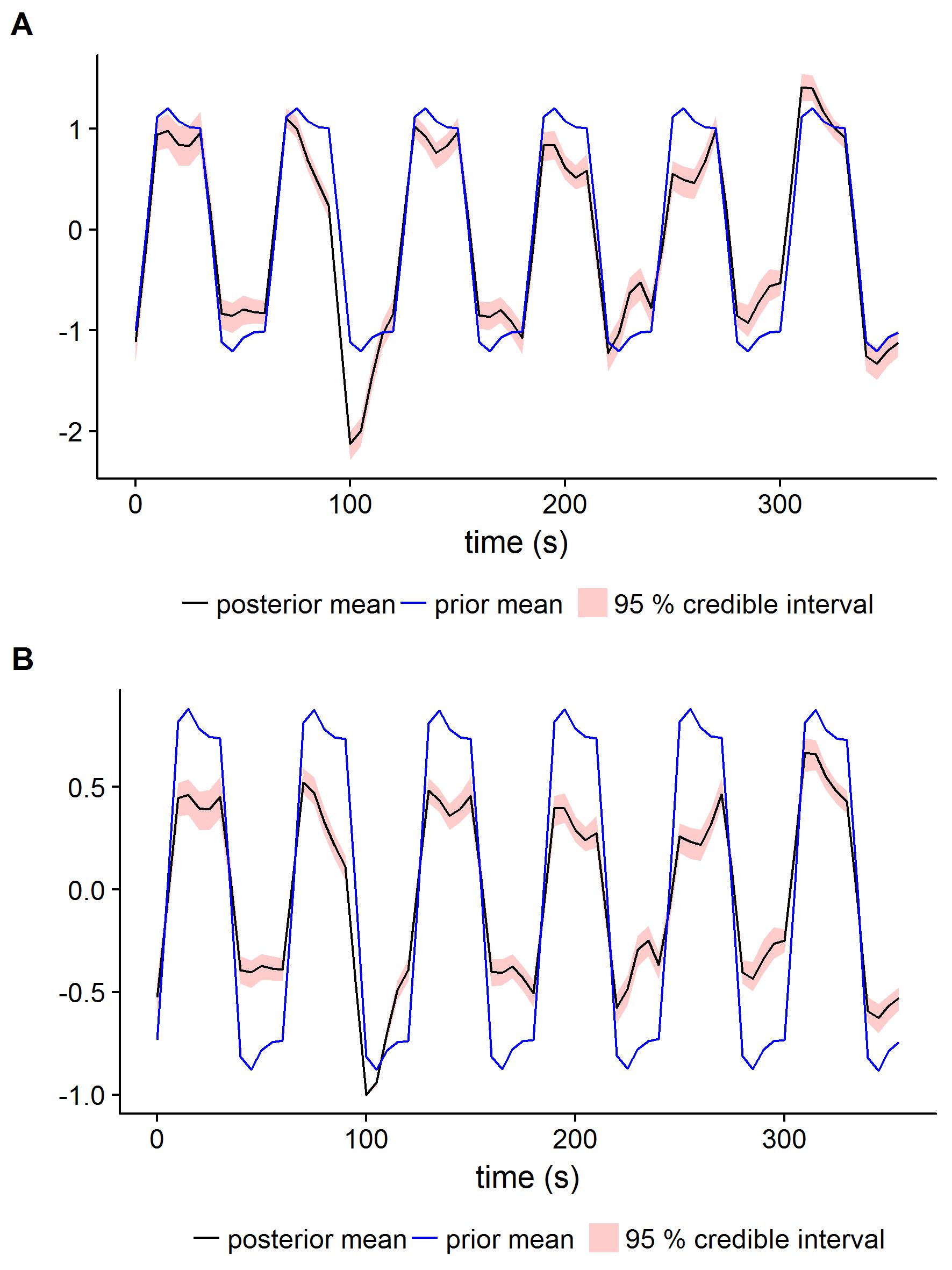}\caption{Estimated predicted BOLD for subject 19628 and parcel 159. (A) is
the Gaussian process $\mathbf{F}$ in (\ref{eq:full multivar model})
and (B) is the transformed Gaussian process $H\left(\mathbf{F}\right)$,
see Equation (\ref{eq:first trans}) and (\ref{eq: trans final}).\label{fig:sub19_pbold_157}}
\end{figure}

\begin{figure}
\includegraphics[scale=0.48]{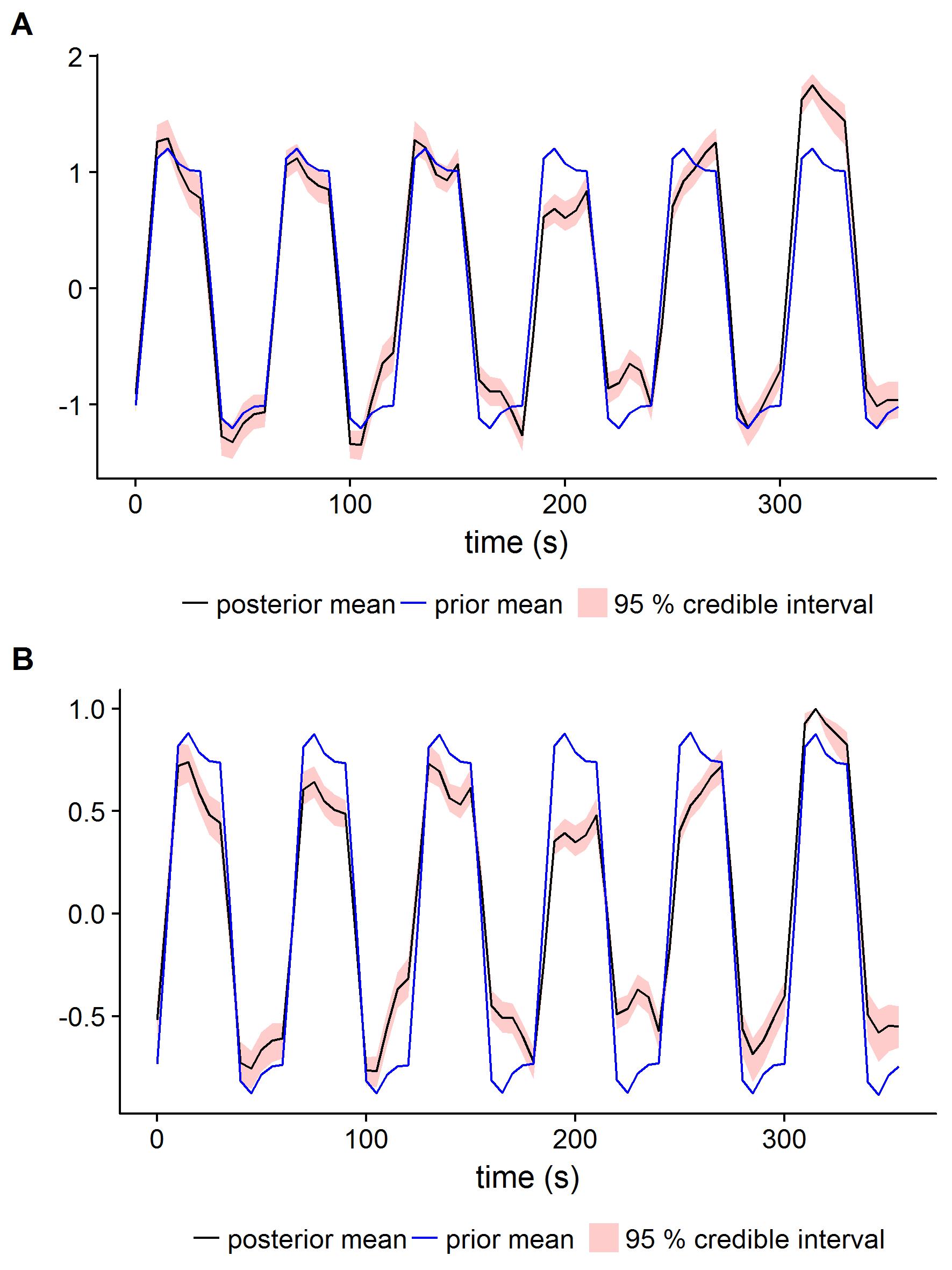}\caption{Estimated predicted BOLD for subject 19628 and parcel 32. (A) is the
Gaussian process $\mathbf{F}$ in (\ref{eq:full multivar model})
and (B) is the transformed Gaussian process $H\left(\mathbf{F}\right)$,
see Equation (\ref{eq:first trans}) and (\ref{eq: trans final}).
\label{fig:sub19_pbold_32}}
\end{figure}

Figures \ref{fig:sub19_pbold_157} and \ref{fig:sub19_pbold_32} show
the estimated posterior for the predicted BOLD in the two parcels
with most positive activation for subject 19628. Parcel 159 is the
yellow cluster in Z-slice 10 and 11 in Figure \ref{fig:t-maps_real_one_basis}
and has 47 active voxels in total. Parcel 32 is the purple cluster
in Z-slice 10 and 11 in Figure \ref{fig:t-maps_real_one_basis} and
has 47 active voxels in total. Note that the scale of the transformed
GP is relative and bounded between -1 and 1, due to the infinity norm.
The posteriors have a non-linear behavior, where the amplitude of
the peaks and the undershoots vary over time. This form of the posteriors
indicate that the data contain important information about the shape
of the predicted BOLD, which is not contained in the prior. 

\begin{figure*}
\centering{}\includegraphics[width=14.7273cm,height=9cm]{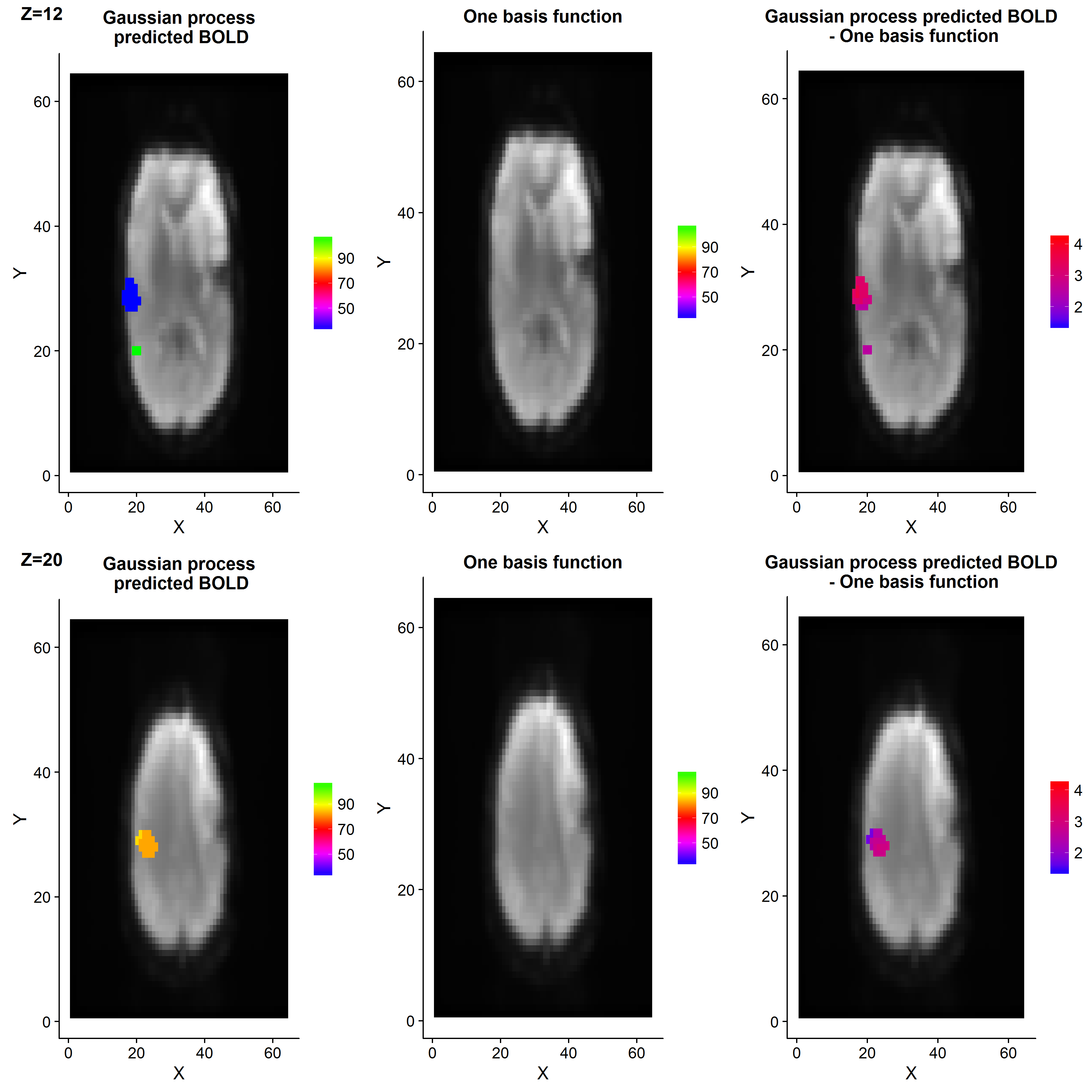}\caption{Example slices with Bayesian $t$-ratios for subject 18716. The activity
maps are thresholded at $t\ge4$ for a test that tests the effect
size $0.25$. The color specifies parcel belonging for active voxels.
The rightmost column shows the differences in t-ratios, thresholded
such that only values fulfilling $|t_{1}-t_{2}|>1$ are shown. Top
row: the flexible model finds brain activity in the auditory cortex,
which is not found using the fix model. Bottom row: the flexible model
finds brain activity in the motor cortex (generated by speech production),
which is not found using the fix model. \label{fig: tmap_sub18_one_basis}}
\end{figure*}

\begin{figure*}
\centering{}\includegraphics[width=14.7273cm,height=9cm]{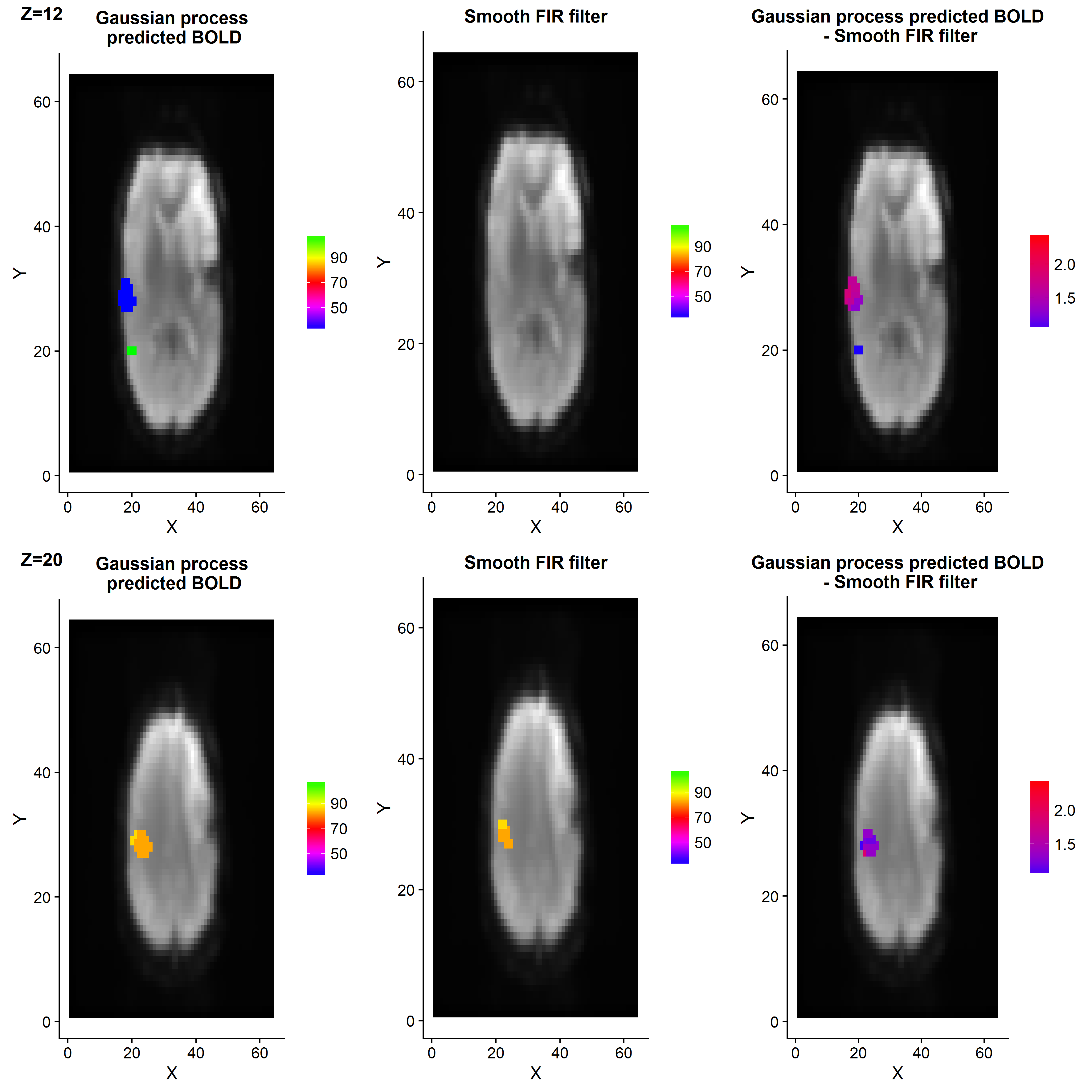}\caption{Example slices with Bayesian $t$-ratios for subject 18716. The activity
maps are thresholded at $t\ge4$ for a test that tests the effect
size $0.25$. The color specifies parcel belonging for active voxels.
The rightmost column shows the differences in t-ratios, thresholded
such that only values fulfilling $|t_{1}-t_{2}|>1$ are shown. \label{fig: tmap_sub18_FIR}}
\end{figure*}

Turning now to subject 18716, none of the two first baseline models
(using standard basis functions) detected any activity for the given
effect size, but our model that estimates the predicted BOLD detected
several active voxels, which can be seen in Figure \ref{fig: tmap_sub18_one_basis}.
For example, the flexible model detects brain activity in auditory
cortex and in motor cortex, not detected by the fix model. As can
be seen in Figure \ref{fig: tmap_sub18_FIR}, the smooth FIR filter
approach also detects activity in the motor cortex, but not in the
auditory cortex. It should be noted, however, that this activity difference
is not caused by a different HR due to the tumor, as the detected
activity is on the opposite side of the tumor. The results for the
baseline model with two basis functions are not shown.

\begin{figure}
\includegraphics[scale=0.48]{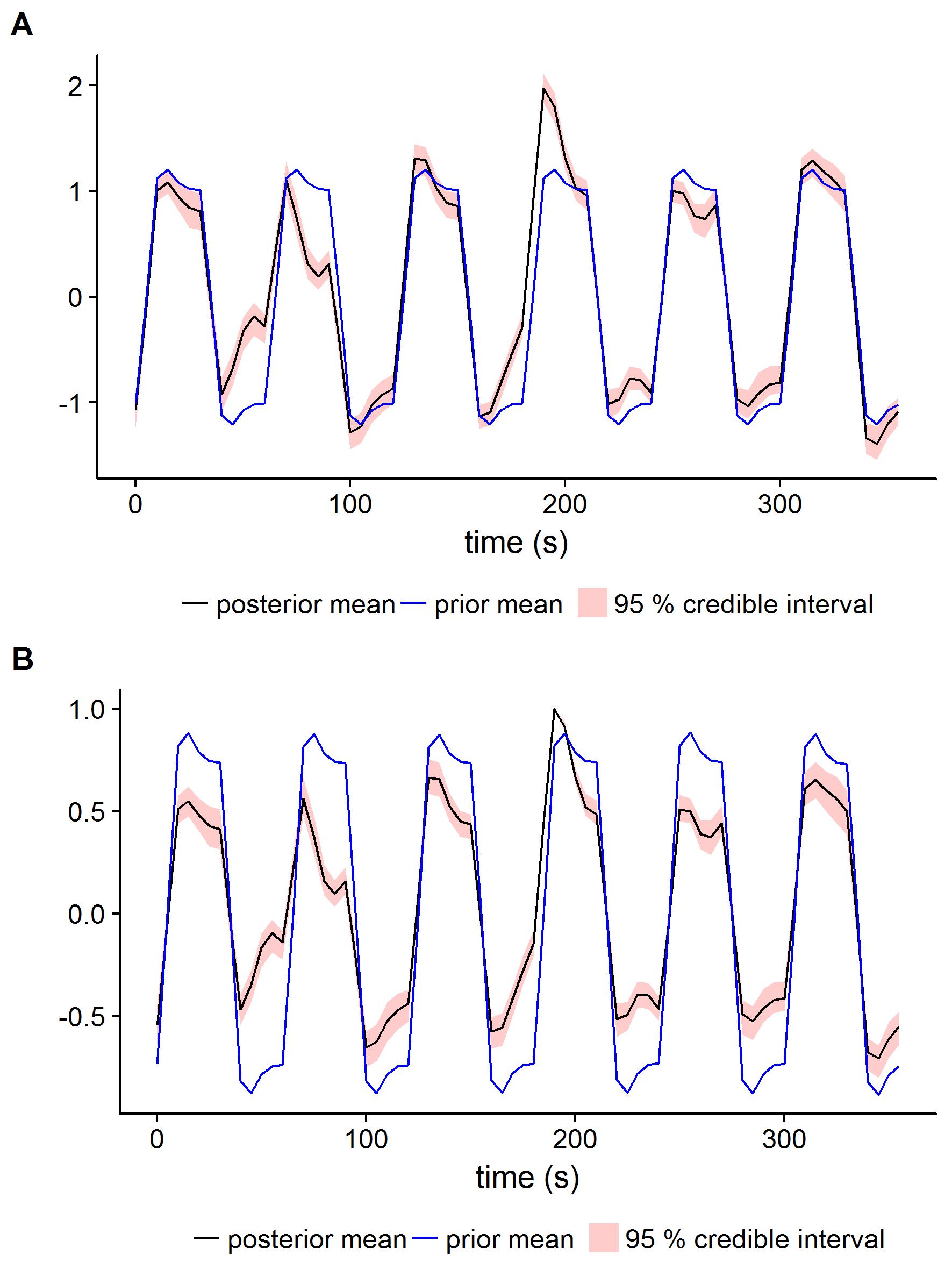}\caption{Estimated predicted BOLD for subject 18716 and parcel 32. (A) is the
Gaussian process $\mathbf{F}$ in (\ref{eq:full multivar model})
and (B) is the transformed Gaussian process $H\left(\mathbf{F}\right)$,
see Equation (\ref{eq:first trans}) and (\ref{eq: trans final}).
\label{fig:sub18_pbold_32}}
\end{figure}

Figure \ref{fig:sub18_pbold_32} shows the estimated posterior for
the predicted BOLD in one of the parcels with most positive activation
for subject 18716. Parcel 32 is the blue cluster in Z-slice 12 in
Figure \ref{fig: tmap_sub18_one_basis} and has 25 active voxels in
total. Similar to the posterior predicted BOLD shown for subject 19626,
the amplitudes of the peaks and the undershoots are non-stationary.

\begin{figure*}
\centering{}\includegraphics[width=15.5455cm,height=9cm]{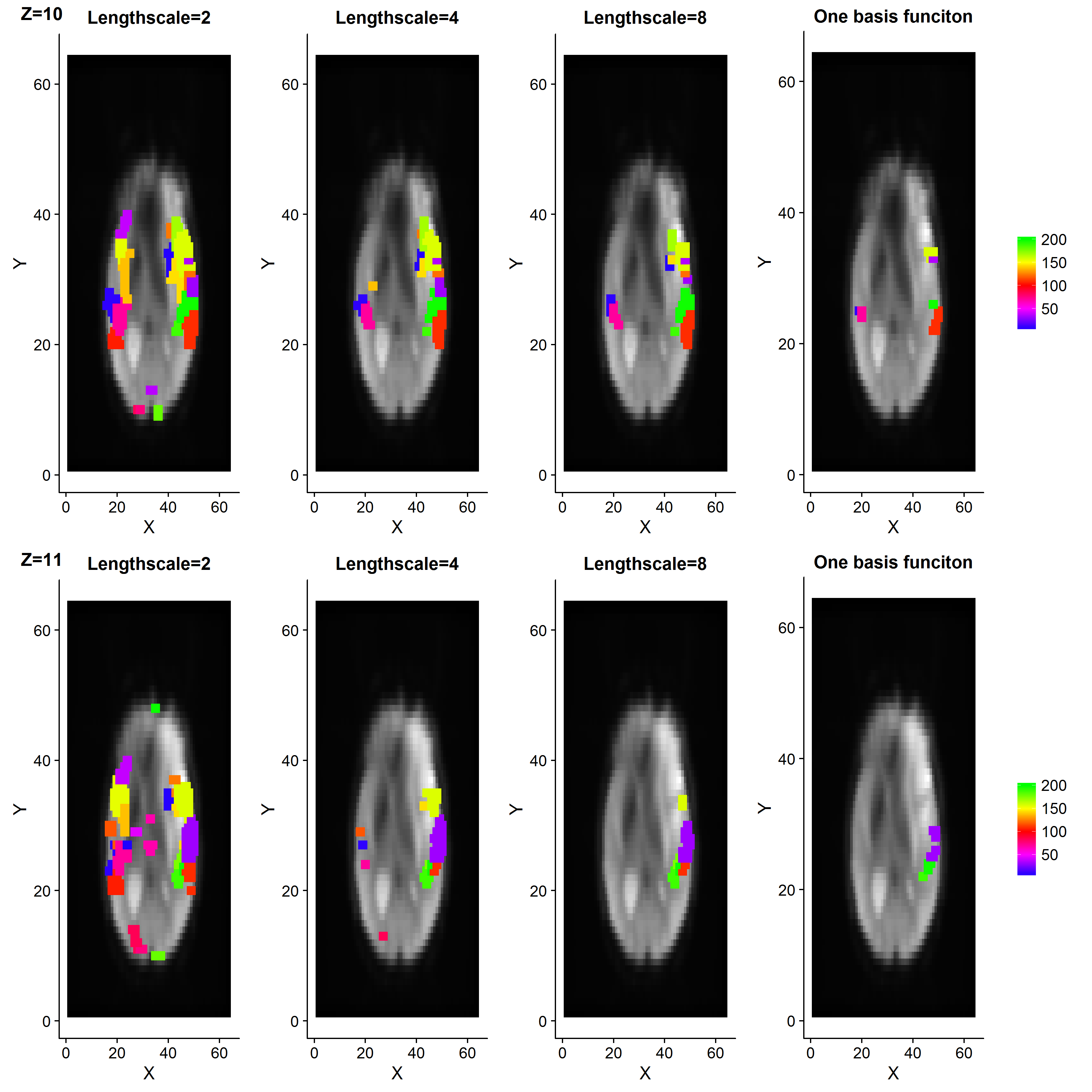}\caption{Comparison of different lengthscales for the GP model. Example slices
with Bayesian $t$-ratios for subject 19628. The activity maps are
thresholded at $t\ge4$ for a test that tests the effect size $0.25$.
The color specifies parcel belonging for active voxels. Clearly, decreasing
the lengthscale leads to a more flexible predicted BOLD, which may
lead to overfitting. \label{fig:t-maps_real_lengthscale}}
\end{figure*}

In order to investigate the effect of the lengthscale hyperparameter,
GP models with different lengthscales were estimated. The result is
presented in Figure \ref{fig:t-maps_real_lengthscale}. With the shorter
lengthscales, the predicted BOLD gets more flexible, and thus finds
more activity. Similarly, longer lengthscales restrict the model and
the results are more similar to the reference model.

In parcels that lack activity, the posterior predicted BOLD is similar
to the prior distribution (not shown).

\begin{figure*}
\begin{centering}
\includegraphics[scale=0.3]{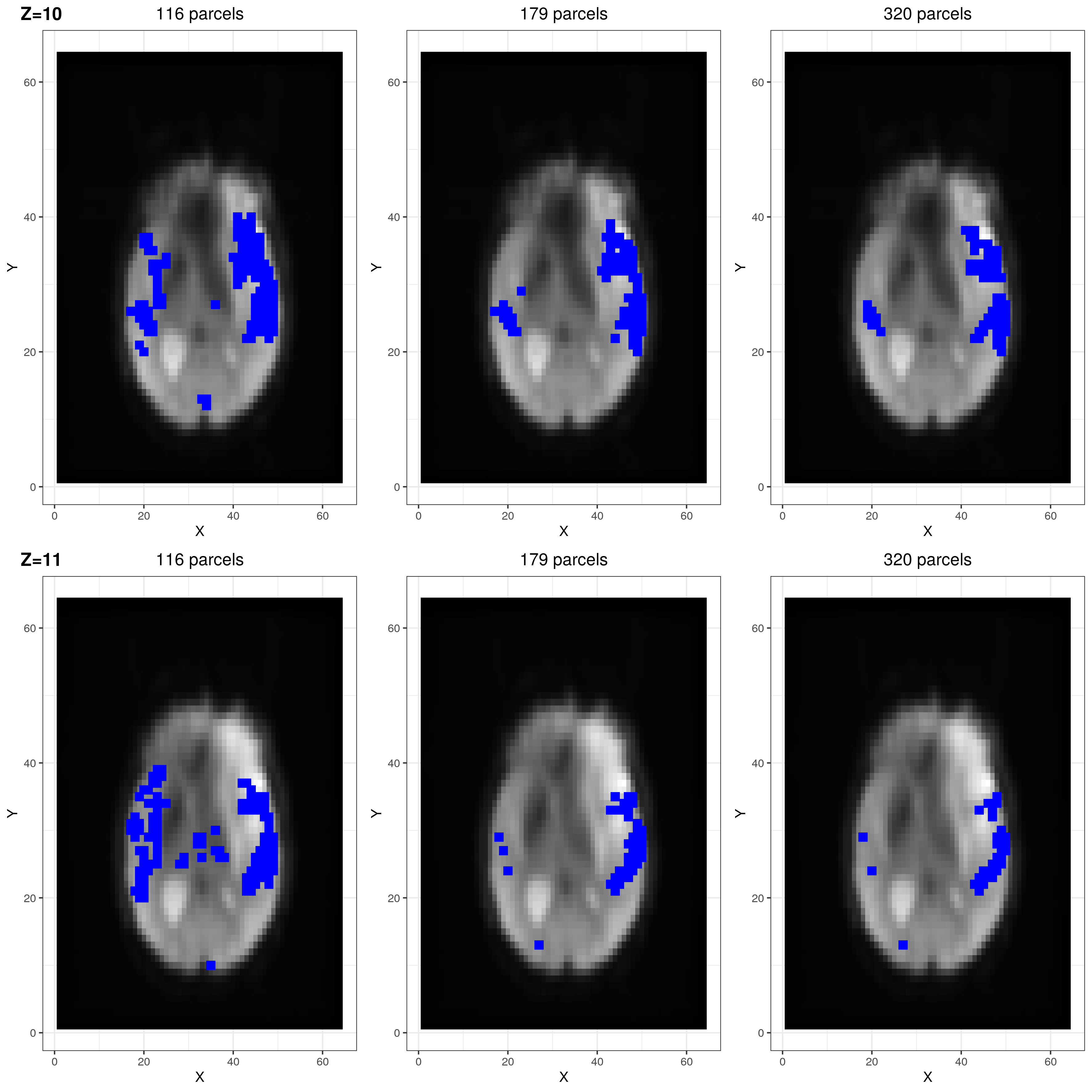}
\par\end{centering}
\caption{Comparing activation patterns across three different parcellations
for two example slices for subject 19628. The middle column displays
the results from the parcellation used for the other results in this
section. The activity maps are thresholded at $t\ge4$ for a test
that tests the effect size $0.25$.\label{fig:ParcelAtlasActivations}}
\end{figure*}

Finally, Figure \ref{fig:ParcelAtlasActivations} and Table \ref{tab:ParcelAtlasFalsePositives}
shows how the activations changes when two two alternative atlases
for parcellations are used. Smaller atlas with 116 parcels is the
Eickhoff-Zilles parcel atlas \citep{eickhoff2005new,craddock2013neuro}.
The larger atlas is the ADHD 400 atlas \citep{craddock,bellec2017neuro};
note that the brain mask removes some parcels. The results are essentially
unchanged when using the ADHD 400 atlas. The atlas with the smallest
number of parcels give substantially more voxels are active for both
subjects. The reason for this is that our model uses the same AR process
for all voxels within a parcel. When the parcels contains many voxels,
this becomes restrictive and the GP finds patterns in the remaining
noise that sometimes happens to correlate with the paradigm. For atlases
with few parcels it is better to use separate AR processes in each
voxel, which is a straightforward extension; see Section \ref{subsec:Future-work}.

\begin{table}
\centering{}%
\begin{tabular}{lcccccc}
\hline 
No. of parcels &  & \multicolumn{2}{c}{Subject 18716} &  & \multicolumn{2}{c}{Subject 19628}\tabularnewline
 &  & GP & Fix &  & GP & Fix\tabularnewline
\cline{3-4} \cline{4-4} \cline{6-7} \cline{7-7} 
\multirow{1}{*}{116} &  & 3.48\% & 0 &  & 8.82\% & 0.15\%\tabularnewline
\multirow{1}{*}{186/179} &  & 0.82\% & 0 &  & 2.41\% & 0.26\%\tabularnewline
\multirow{1}{*}{336/320} &  & 0.40\% & 0 &  & 1.36\% & 0.19\%\tabularnewline
\hline 
\end{tabular}\caption{Percentage of active voxels for different parcellations. The number
of parcels varies since the brain mask removes some parcels.\label{tab:ParcelAtlasFalsePositives}}
\end{table}

\section{Discussion}

\subsection{Model the predicted BOLD instead of the HRF}

We have proposed and implemented a new way to model the predicted
BOLD for task fMRI. The difference from other models is the direct
modeling of the predicted BOLD response, instead of the HRF, combined
with a straight forward measure of activity. The simulation study
shows that the model has a good ability to discriminate between active
and non-active voxels. The proposed model shows robustness to misspecification
in the prior mean function for the predicted BOLD. This is a desirable
feature, since it is likely that a model with a fix predicted BOLD
will not be correct for the whole brain or across subjects. Our proposed
model gives the researcher a framework to approach problems related
to the HR. For example, for group studies (where all data are transformed
to a standard space), the predicted BOLD in one parcel can be compared
across subjects. Also, the existence of explicit activity parameters
makes it easy to construct PPMs or $t$-maps.

The non-linear aspect of the hemodynamics can be captured with the
GP model. It is interesting to study the properties of the posterior
for the predicted BOLD, see Figures \ref{fig:sub19_pbold_157}, \ref{fig:sub19_pbold_32}
and \ref{fig:sub18_pbold_32}. Compared to the prior mean function,
the major difference for the posterior is the time-varying amplitude
of the peaks and the undershoots. This feature seems to be crucial
to find the additional activity compared to the two baseline models
and the smooth FIR approach. The described feature is not easily incorporated
into traditional GLM approaches. Parametric modulation of the HRF
can be used in the LTI context, to obtain hemodynamic features that
are non-stationary, but this approach comes with two problems. First,
a proper modulator must be chosen. Second, the non-stationarity is
assumed to be be known and fix across the brain given the modulator.
Our approach handles the non-stationarity in an unsupervised manner,
and can of course use a prior mean function that depends on a problem
specific modulator.

\subsection{Computation}

The model is implemented in the R programming language, and the Markov
Chain Monte Carlo (MCMC) code is not optimized for speed, except for
the CPU parallelization of models across parcels. The computations
for our model are therefore rather time consuming, but there are several
options to reduce the computational time. One option is to use other
inference methods, such as VB inference. Another option is to use
GPUs (graphics cards) for parallelization \citep{eklund2012fmri,eklund2014broccoli},
particularly if there are a relatively large number of parcels. Another
way to reduce the processing time is to stop the calculations for
a given parcel, if it after a certain number of draws is unlikely
that even a single voxel in the parcel will survive an activation
threshold.

\subsection{Multiple comparisons}

In contrast to frequentist methods, there is no consensus in the fMRI
field regarding if and how to correct for multiple comparisons for
PPMs. In frequentist hypothesis testing the null hypothesis is normally
that the parameter representing the brain activity is 0, but using
an effect size threshold of 0 for PPMs often leads to activation in
a very large portion of the voxels (even for strict probability thresholds
for the PPMs). In this paper we have mainly focused on differences
between fix and flexible predicted BOLD models, using voxel inference
and an effect size threshold of 0.25. One ad hoc approach to correct
for multiple comparisons is to calculate a Bayesian $t$- or $z$-score
for each voxel, and then apply existing frequentistic approaches for
multiple comparison correction (e.g. Gaussian random field theory).
This approach is for example used in the FSL software.

\subsection{Applications}

There are several possible applications of our proposed model. As
demonstrated in this paper, a potential application is in clinical
fMRI, where fMRI can be used to map out important brain areas prior
to tumor surgery. The HR may be different close to a tumor, and our
flexible model can then be used to detect more brain activity, and
thereby potentially lead to a better treatment plan. Other cases where
the HR may be different include young subjects \citep{richter2003shape},
subjects with epilepsy \citep{jacobs2008variability} and subjects
with stroke \citep{bonakdarpour2007hemodynamic}. Our model can also
be used to automatically handle cases where a subject fails to perform
one or several events, or where a subject occasionally struggles with
the timing of the experiment (adding a temporal derivative can only
account for a global shift in time, and not local time shifts). As
mentioned in the introduction, our model can also pick up variations
in the strength of the BOLD response, while virtually all other models
see the stimulus as a fixed effect \citep{westfall2016fixing}.

\subsection{Future work\label{subsec:Future-work}}

A natural extension of the model is to do inference for the GP hyperparameters.
Since the model uses a non-Gaussian likelihood this is, however, non-trivial.
In the MCMC case the methods presented in \citep{filippone2014pseudo,murray2016pseudo}
could be used, where pseudo marginal inference is employed together
with an unbiased estimate of the intractable marginal likelihood for
the GP. 

The prior mean is a quite severely misspecified in the simulation
in Section \ref{sec:Simulations}, and the GP model cannot fully capture
the true underlying signal, see Figures \ref{fig:sim_pbold_l=00003D4}
and \ref{fig:sim_pbold_l=00003D2}, even though it captures sufficient
variation to be able to distinguish between active and non-active
voxels. Other kernels could perform even better in this scenario.
For block paradigms, locally periodic kernels are a natural candidate,
see for example \citep{duvenaud2014automatic} for a discussion. Those
kernels could introduce a positive periodic correlation that decays
with the distance; paradigm blocks near in time will be more correlated
compared to blocks further away in time. Another alternative is use
the prior mean function $m\left(t\right)$ and the derivative $\partial m\left(t\right)/\partial t$
as covariates in a smooth kernel. This would give a behavior similar
to locally periodic kernel, but can also handle event related data.
The LTI system approximation of the hemodynamics works fairly well
in many scenarios. The GP on the FIR filter is a efficient parameterization
for systems that is linear or close to linear. One approach to exploit
the advantages of the FIR kernel is to construct a kernel for the
predicted BOLD that is a linear combination of a FIR kernel and a
global smooth kernel that will account for non-linear effects. The
number of hyperparameters grows with more complex kernels, and it
is hard to manually specify these parameters. To efficiently use more
complex kernels, inference for the hyperparameters is important, either
using an estimate of the posterior mode or sampling via MCMC. Even
more sophisticated kernels could be used, such as spectral mixture
kernels \citep{wilson2013gaussian} or deep kernels \citep{wilson2016deep}.
Such kernels are very expressive and can find complicated patterns
given sufficient data. However, the proposed inference methods are
adapted for the classical GP regression model $y=f\left(x\right)+\epsilon$,
where $f$ is a GP, and not as a part of a multivariate time series
regression model. The problem boils down to finding \textit{data driven}
features that can be used in some kernel, in order to explain the
prior correlation in $\mathbf{f}$ in a good way for a given parcel.

Another possible future direction is to use more sophisticated priors
on $\mathbf{B}$ and $\boldsymbol{\Gamma}$. Here variable selection
would be appropriate, as it could reduce overfitting and improve the
model fit. Interesting selection methods could be spike and slab priors
or horseshoe priors. Due to the spatial nature of the data, spatial
priors would also be an interesting choice. For example, a GP prior
could be placed over $\mathbf{B}$ to impose a spatial smoothness,
e.g. using a Matérn kernel. This would result in models similar to
those in \citep{luttinen2009variational} and \citep{wilson12icml}.
Sidén et al. \citep{siden2017fast} use sparse precision matrices
to model spatial dependencies in whole brain task fMRI data, and derive
both fast MCMC and VB methods. Those ideas could be incorporated into
the proposed model.

The results in Table \ref{tab:ParcelAtlasFalsePositives} indicates
that inferred activations is robust against parcellation atlas, unless
the number of parcels is very small. Since the model assumes the same
hemodynamics for all voxels within a parcel, an interesting direction
for future work is to learn the optimal parcellation dynamically from
data \citep{chaari2012hemodynamic,chaari:hal-01255465,albughdadi2017}
using a more sophisticated regularization of the Gaussian process
to avoid overfitting. For atlases with a small number of parcels,
it makes sense to use voxel-wise AR processes for the noise, and this
is a straightforward extension.

The suggested model is made for single subject data, and in many cases
joint inference for many subjects is desirable. A hierarchical model
could be used here, with random effects for the $\mathbf{B}$ and
$\mathbf{F}$ parameters. Assume that all subjects have been transformed
to the same space with the same parcellation. The prior for $\mathbf{f}_{m}$
for a given parcel could then be expressed as

\[
\begin{array}{c}
\mathbf{f}_{m}\sim N\left(m\left(t\right),\mathtt{\mathtt{k}}\left(t,t'\right)\right)\\
\mathbf{f}_{m,n}\sim N\left(\mathbf{f}_{m},\varsigma_{m}\mathbb{I}\right)
\end{array},
\]
where $n$ is subject index and $\varsigma_{m}$ is the random effects
variance. $\mathbf{B}$ can be modeled in the same way. This construction
is similar to the within subject models used in \citep{chaari2012hemodynamic,chaari:hal-01255465,albughdadi2017},
but a difference is that they have a random effect on the HRF filter
and the hierarchy is over parcels and voxels.

\section{Conclusion}

We have proposed a novel framework for modeling the hemodynamics in
task fMRI. The new model is shown to more accurately detect brain
activity compared to traditional parametric and nonparametric LTI
models. We model the predicted BOLD directly with a GP prior, as a
part of larger time series regression model. We also introduce an
identifying transformation that solves the challenging identification
problem present in bilinear models in the JDE context. Our new framework
gives researchers the opportunity to ask new kinds of questions related
to hemodynamics, especially with regard to non-linear effects.

\section*{Acknowledgments}

This work was funded by Swedish Research Council (Vetenskapsrådet)
grant no. 2013-5229. Anders Eklund was funded by Center for Industrial
Information Technology (CENIIT) at Linköping University. Declarations
of interest: none

\section{References}

\bibliographystyle{plainnat}
\bibliography{paper_arxiv1}

\section*{Appendix A: Distributions\label{sec:AppendixA}}

The density function for a matrix normal distribution

\[
\mathbf{Y}\sim MN_{n\times p}\left(\mathbf{M},\mathbf{U},\mathbf{V}\right)
\]

is of the form

\[
p\left(\mathbf{Y}|\mathbf{M},\mathbf{U,}\mathbf{V}\right)=\frac{exp\left(-\frac{1}{2}tr\left[\mathbf{V}^{-1}\left(\mathbf{Y}-\mathbf{M}\right)^{\top}\mathbf{U}^{-1}\left(\mathbf{Y}-\mathbf{M}\right)\right]\right)}{\left(2\pi\right)^{np/2}|\mathbf{V}|^{n/2}|\mathbf{U}|^{p/2}}
\]

Parameters:
\begin{itemize}
\item $\mathbf{M}$: location, real $n\times p$ matrix
\item $\mathbf{U}$: scale, positive-definite real $n\times n$ matrix (dependencies
over observations)
\item $\mathbf{V}$: scale, positive-definite real $p\times p$ matrix (dependencies
over variables)
\end{itemize}
The matrix normal distribution is related to the multivariate normal
distribution in the following way:

\[
vec\left(\mathbf{Y}\right)\sim N\left(vec\left(\mathbf{M}\right),\mathbf{V}\otimes\mathbf{U}\right)
\]

\section*{Appendix B: Likelihood function and priors\label{subsec:Parameters-and-priors}}

The likelihood function for the model in (\ref{eq:full multivar model})
is of the form

\begin{multline*}
L\left(\mathbf{Y}|\mathbf{F},\mathbf{B},\boldsymbol{\Gamma},\boldsymbol{\sigma}^{2},\boldsymbol{\rho},\mathbf{Z}\right)=\left(2\pi\right)^{-TJ/2}|\boldsymbol{\Omega}|^{-T/2}|\mathbf{M}_{\boldsymbol{\rho}}|^{-J/2}\times\\
\exp\left(-\frac{1}{2}\mathrm{tr}\left[\boldsymbol{\Omega}^{-1}\mathbf{\bar{Y}}^{\top}\mathbf{M}_{\mathbf{\boldsymbol{\rho}}}^{-1}\mathbf{\bar{Y}}\right]\right),
\end{multline*}
where $\mathbf{\bar{Y}}=\mathbf{Y}-H\left(\mathbf{F}\right)\mathbf{B}-\mathbf{Z}\boldsymbol{\Gamma}$.

The model (\ref{eq:full multivar model}) has the following parameters
and priors:
\begin{enumerate}
\item $\mathbf{F}$ has independent Gaussian process priors on each column,
i.e.
\[
\mathbf{f}_{m}\sim N\left(\mathbf{f}_{0,m},\mathtt{K}\left(\mathcal{T_{\star}},\mathcal{T_{\star}}\right){}_{m}\right)\quad m=1,\ldots,M,
\]
where $\mathbf{f}_{0,m}$ is the mean function $m\left(t\right)$
evaluated at the time points $\mathcal{T}_{\star}$.
\item The elements of $\boldsymbol{\sigma}^{2}$ are assumed to be independent
a priori and are modeled as
\[
\begin{array}{c}
\sigma_{j}^{2}\sim\mathrm{InvGamma}(c_{0,j},d_{0,j}).\end{array}
\]
\item We use the standard conjugate priors for multivariate regression \citep{press1982}
for the regression parameters $\mathbf{B}$ and $\mathbf{\boldsymbol{\Gamma}}$.
The prior for $\mathbf{B}$ is modeled conditional on $\boldsymbol{\Omega}=\mathrm{diag}(\boldsymbol{\sigma}^{2})$
as a matrix normal distribution
\[
\begin{array}{c}
\mathbf{B}|\boldsymbol{\Omega}\sim MN_{M\text{\ensuremath{\times}J}}\left(\mathbf{B}_{0},\boldsymbol{\Omega},\kappa^{-1}\mathbf{P}^{-1}\right),\end{array}
\]
where $\mathbf{P}$ is a $M\times M$ positive definite precision
matrix over stimuli, $\mathbf{B}_{0}$ is the $M\times J$ prior mean
matrix and $\kappa$ is a scalar. $\mathbf{\boldsymbol{\Gamma}}$
is assigned a matrix normal prior conditional on $\boldsymbol{\Omega}$
\[
\begin{array}{c}
\boldsymbol{\Gamma}|\boldsymbol{\Omega}\sim MN_{M\text{\ensuremath{\times}J}}\left(\boldsymbol{\Gamma}_{0},\boldsymbol{\Omega},\tau^{-1}\mathbb{I}_{P}\right).\end{array}
\]
\item Following \citep{eklund2017}, the prior on the AR process parameters
is centered over a stationary AR(1) process: 
\[
\begin{array}{c}
\boldsymbol{\rho}\sim N\left(\boldsymbol{\rho}_{0},\mathbf{A}_{0}\right)\cdot I\left(\boldsymbol{\rho}\right),\end{array}
\]
where $\boldsymbol{\rho}_{0}=(r,0,\ldots,0)$, $\mathbf{A}_{0}=\mathrm{diag}(c^{2},\frac{c^{2}}{2^{\zeta}},\ldots,\frac{c^{2}}{K^{\zeta}})$
and $I\left(\boldsymbol{\rho}\right)$ is an indicator function for
the stationary region
\[
I\left(\boldsymbol{\rho}\right)=\begin{cases}
1 & \text{ if }\left|\text{\ensuremath{\ell}}_{max}\right|<1\\
0 & \text{ otherwise, }
\end{cases}
\]
where $\ell_{max}$ is the largest absolute (modulus) eigenvalue of
the companion matrix
\[
\left(\begin{array}{ccccc}
\rho_{1} & \rho_{2} & \cdots & \rho_{K-1} & \rho_{K}\\
1 & 0 &  & 0 & 0\\
0 & 1 &  &  & 0\\
 &  & \ddots & 0 & \vdots\\
0 &  & 0 & 1 & 0
\end{array}\right).
\]
Note that the prior is centered over the noise process $u_{t}=\rho_{1}\cdot u_{t-1}+\epsilon_{t}$,
but assigns probability mass also to higher order AR processes in
such a way that longer lags are shrunk more heavily toward zero.
\end{enumerate}
The following prior hyperparameters are used in the paper, unless
otherwise noted:
\begin{enumerate}
\item $\mathbf{F}$: We use two different lengthscales: $l=2$ or $l=4$,
and $\omega=\sqrt{0.1}\approx0.316$.
\item $\boldsymbol{\sigma}^{2}$: $c_{0}=0$ and $d_{0}=0$, giving a non-informative
prior.
\item $\mathbf{B}$: precision scale factor $\kappa=10^{-10}$, precision
matrix $\mathbf{P}$ is diagonal, giving an essentially flat non-informative
prior.
\item $\mathbf{\boldsymbol{\Gamma}}$: $\tau=0$, except for the parameter
representing the constant, which was given an empirical prior based
on voxel mean and four times the voxel variance for each voxel. The
nuisance variables are scaled to have zero mean and unit variance.
\item $\boldsymbol{\rho}$: $\boldsymbol{\rho}_{0}=(0,0,0)$, $\mathbf{A}_{0}=\mathrm{diag}(0.5,\frac{0.5}{2^{5}},\frac{0.5}{3^{5}})$.
\end{enumerate}

\section*{Appendix C: Gibbs sampling\label{app:PosteriorComputations}}

\subsection*{Sampling $\boldsymbol{\rho}$}

The autoregressive parameters are sampled using the following formulation

\[
\mathbf{u}=\rho_{1}\mathbf{u}_{-1}+\rho_{2}\mathbf{u}_{-2}+\ldots+\rho_{k}\mathbf{u}_{-k}+\boldsymbol{\epsilon}\Longleftrightarrow
\]

\begin{equation}
\mathbf{u}=\left(\begin{array}{ccc}
\mathbf{u}_{-1} & \cdots & \mathbf{u}_{-k}\end{array}\right)\left(\begin{array}{c}
\rho_{1}\\
\vdots\\
\rho_{k}
\end{array}\right)+\boldsymbol{\epsilon}=\mathbf{D}\boldsymbol{\rho}+\boldsymbol{\epsilon},\label{eq:u regression form}
\end{equation}
where $\boldsymbol{u}_{\star}=vec\left(\mathbf{Y}_{\star}-\mathbf{F}_{\star}\mathbf{B}-\mathbf{Z}_{\star}\boldsymbol{\Gamma}\right)$
is used to calculate $\mathbf{u}$, given that $\mathbf{B}$, $\mathbf{F}_{\star}$
and $\boldsymbol{\Gamma}$ are known, $\mathbf{D}$ is a matrix of
size $JT\times k$. \textbf{$\mathbf{u}$} and $\mathbf{D}$ must
be updated in every iteration since $\mathbf{B}$, $\mathbf{F}$ and
$\boldsymbol{\Gamma}$ also are updated in every iteration. Let $\text{\ensuremath{\boldsymbol{\Sigma}}}=\boldsymbol{\Omega}\otimes\mathbb{I}_{T}$,
then $\boldsymbol{\epsilon}\sim N\left(\mathbf{0},\text{\ensuremath{\boldsymbol{\Sigma}}}\right)$.
It is clear from (\ref{eq:u regression form}) that the full conditional
posterior for $\boldsymbol{\rho}$ can be obtained using standard
formulas for univariate regression with heteroscedastic variance,
and the full conditional posterior is given by

\begin{equation}
\begin{array}{c}
\boldsymbol{\rho}_{n}|\boldsymbol{\Sigma}\sim N\left(\boldsymbol{\rho}_{n},\mathbf{A}_{n}\right)\\
\boldsymbol{\rho}_{n}=\mathbf{A}_{n}\left(\mathbf{D}^{\top}\text{\ensuremath{\boldsymbol{\Sigma}}}^{-1}\mathbf{D}\hat{\boldsymbol{\rho}}+\mathbf{A}_{0}\boldsymbol{\rho}_{0}\right)\\
\mathbf{A}_{n}=\left(\mathbf{D}^{\top}\text{\ensuremath{\boldsymbol{\Sigma}}}^{-1}\mathbf{D}+\mathbf{A}_{0}^{-1}\right)^{-1}\\
\hat{\boldsymbol{\rho}}=\left(\mathbf{D}^{\top}\text{\ensuremath{\boldsymbol{\Sigma}}}^{-1}\mathbf{D}\right)^{-1}\mathbf{D}^{\top}\text{\text{\ensuremath{\boldsymbol{\Sigma}}}}^{-1}\mathbf{u}.
\end{array}\label{eq:rho_full_cond_post}
\end{equation}
To make sure that $\boldsymbol{\rho}$ is in the stationary region,
draws from $\boldsymbol{\rho}_{n}|\boldsymbol{\Sigma}\sim N\left(\boldsymbol{\rho}_{n},\mathbf{A}_{n}\right)$
are discarded until a draw that is inside the stationary region is
obtained. 

\subsection*{Sampling $\mathbf{B}$, $\boldsymbol{\Gamma}$ and $\boldsymbol{\sigma}^{2}$}

In order to simplify sampling of $\mathbf{B}$ and $\boldsymbol{\Gamma}$
in model (\ref{eq:full multivar model}) the following formulation
is used

\[
\begin{array}{c}
\mathbf{Y}_{\star}=H\left(\mathbf{F}\right)\mathbf{B}+\mathbf{Z}_{\star}\boldsymbol{\Gamma}+\mathbf{U}_{\star}\Leftrightarrow\\
\mathbf{Y}_{\star}=\left(\begin{array}{cc}
H\left(\mathbf{F}\right) & \mathbf{Z}_{\star}\end{array}\right)\left(\begin{array}{c}
\mathbf{B}\\
\boldsymbol{\Gamma}
\end{array}\right)+\mathbf{U}_{\star}
\end{array}
\]

\begin{equation}
\mathbf{Y}_{\star}=\mathbf{X}_{\star}\mathbf{Q}+\mathbf{U}_{\star},\label{eq:mult-reg B Gamma}
\end{equation}
where $\mathbf{X}_{\star}$ is of the size $T_{\star}\times\left(M+P\right)$
and $\mathbf{Q}$ is of the size $\left(M+P\right)\times J$. In order
to use the standard multivariate regression formulas in (\ref{eq:mult-reg B Gamma}),
pre-whitening is used. Let $\Phi_{C}(L)$ be the column-wise lag polynomial
from time series analysis, i.e. $\Phi_{C}(L)=1-\rho_{1}L-\rho_{2}L^{2}-\ldots-\rho_{k}L^{k}$.
The first $K$ constant observations in \textbf{$\mathbf{Y}_{0}$},
$\mathbf{F}_{0}$ and $\mathbf{Z}_{0}$ are used for obtaining the
first $K$ observations of $\mathcal{T}$. This results in a new regression
formulation

\begin{equation}
\mathbf{\tilde{Y}}=\mathbf{\tilde{X}}\mathbf{Q}+\mathbf{E},\label{eq:lag full model}
\end{equation}
where $\mathbf{\tilde{Y}}=\Phi_{C}(L)\mathbf{Y}$, $\mathbf{\tilde{X}}=\Phi_{C}(L)\mathbf{X}$
and $\mathbf{E=\tilde{U}}=\Phi_{C}(L)\mathbf{U}$.

Given the parcel constant parameters $\mathbf{F}$, $\boldsymbol{\rho}$,
and the regression parameters $\mathbf{Q}$, the likelihood $p\left(\mathbf{\tilde{\mathbf{Y}}},\mathbf{\tilde{X}}|\boldsymbol{\sigma}^{2},\mathbf{Q}\right)$
is independent over voxels, which implies that inference for each
element in $\boldsymbol{\sigma}^{2}$ is performed by regressions
of the form: $\mathbf{\tilde{y}}_{j}=\mathbf{\tilde{X}}\mathbf{q}_{j}+\boldsymbol{\epsilon}_{j}$
, where $\mathbf{\tilde{y}}_{j}$ is the j:th column of $\tilde{\mathbf{Y}}$
and $\mathbf{q}_{j}$ is the j:th column of $\mathbf{Q}$. The full
conditional posterior for $\sigma_{j}^{2}$ is an Inverse-gamma distribution,
which is easily obtained from standard formulas for univariate regression

\begin{equation}
p\left(\sigma_{j}^{2}|\mathbf{\tilde{y}}_{j},\mathbf{\tilde{X}},\mathbf{Q}\right)\sim\mathrm{InvGamma}\left(c_{n},d_{j}\right)\quad j=1,\ldots,J,\label{eq:posterior sigma-1}
\end{equation}

where

\[
\begin{array}{c}
c_{n}=c_{0}+T/2\\
d_{j}=d_{0}+\frac{1}{2}\left(\tilde{\mathbf{y}_{j}}-\tilde{\mathbf{X}}\mathbf{q}_{j}\right)^{\top}\left(\tilde{\mathbf{y}_{j}}-\tilde{\mathbf{X}}\mathbf{q}_{j}\right)
\end{array}\quad j=1,\ldots,J
\]
Using standard formulas for multivariate regression with conjugate
priors, the full conditional posterior for $vec\left(\mathbf{Q}\right)$
is a multivariate normal distribution. The likelihood function for
(\ref{eq:lag full model}) is described by \citet{press1982}. Let
\\
$\hat{\mathbf{Q}}=\left(\mathbf{\tilde{X}}^{\top}\mathbf{\tilde{X}}\right)^{-1}\mathbf{\tilde{X}}^{\top}\tilde{\mathbf{Y}}$,
$\mathbf{S}=\left(\mathbf{\tilde{Y}}-\mathbf{\tilde{X}}\hat{\mathbf{Q}}\right)^{\top}\left(\mathbf{\tilde{\mathbf{Y}}}-\mathbf{\mathbf{\tilde{X}}}\hat{\mathbf{Q}}\right)/T$
and $vec\left(\mathbf{Q}\right)=\boldsymbol{q}$ . Using standard
manipulations, the likelihood can be written as

\[
\begin{array}{c}
p\left(\mathbf{\tilde{Y}}|\mathbf{\tilde{X}},\mathbf{Q},\text{\ensuremath{\boldsymbol{\Omega}}}\right)\propto\\
|2\pi\text{\ensuremath{\boldsymbol{\Omega}}}|^{-T/2}exp\left(-\frac{1}{2}tr\text{\ensuremath{\boldsymbol{\Omega}}}^{-1}T\cdot\boldsymbol{S}\right)\cdot\\
exp\left\{ -\frac{1}{2}\left(\mathbf{q}-\mathbf{\hat{q}}\right)^{\top}\left[\text{\ensuremath{\boldsymbol{\Omega}}}^{-1}\otimes\left(\mathbf{\tilde{X}}^{\top}\mathbf{\mathbf{\tilde{X}}}\right)\right]\left(\mathbf{q}-\mathbf{\hat{q}}\right)\right\} .
\end{array}
\]
The prior for $\mathbf{Q}$ is now expressed as

\[
\begin{array}{c}
\mathbf{Q}|\boldsymbol{\Omega}\sim MN_{M+p,J}\left(\mathbf{Q}_{0},\boldsymbol{\Omega}\otimes\mathbf{P}_{Q}^{-1}\right)\\
\mathbf{Q}_{0}=\left(\begin{array}{c}
\mathbf{B}_{0}\\
\boldsymbol{\Gamma}_{0}
\end{array}\right)\quad\mathbf{P}_{Q}=\left(\begin{array}{cc}
\kappa\mathbf{P} & \mathbf{0}\\
\mathbf{0} & \tau\mathbb{I}
\end{array}\right),
\end{array}
\]
where $\mathbf{Q}_{0}$ has the same size as $\mathbf{Q}$ and $\mathbf{P}_{Q}$
has size $\left(M+P\right)\times\left(M+P\right)$. Using standard
formulas for multivariate regression, the full conditional posterior
for $\boldsymbol{Q}$ is then given by

\begin{equation}
\boldsymbol{q}|\boldsymbol{\Omega},\tilde{\mathbf{Y}},\mathbf{\tilde{X}}\sim N\left[\bar{\mathbf{q}},\boldsymbol{\Omega}\otimes\left(\mathbf{P}_{Q}+\mathbf{\tilde{X}}^{\top}\mathbf{\tilde{X}}\right)^{-1}\right],\label{eq: posterior B conditional omega-1}
\end{equation}
where

\[
\bar{\mathbf{q}}=\left[\boldsymbol{\Omega}\otimes\left(\mathbf{P}_{Q}+\mathbf{\tilde{X}}^{\top}\mathbf{\tilde{X}}\right)^{-1}\right]vec\left[\left(\mathbf{\tilde{X}}^{\top}\tilde{\mathbf{Y}}+\mathbf{P}_{Q}\mathbf{Q}_{0}\right)\boldsymbol{\Omega}^{-1}\right].
\]

\subsection*{Sampling $\mathbf{F}$}

To simplify the sampling of $\mathbf{F}$, we reformulate the model
as

\begin{align}
\mathbf{g}_{\star} & =\mathrm{vec}(\mathbf{Y}_{\star}-\mathbf{Z}_{\star}\mathbf{\boldsymbol{\Gamma}})\nonumber \\
 & =\mathrm{vec}(\mathbf{\mathbf{\mathbb{I}}}_{T_{\star}}H\left(\mathbf{F}\right)\mathbf{B})+vec(\mathbf{\mathbf{U}}_{\star})\nonumber \\
 & =\mathbf{W}\mathbf{f}_{H}+\boldsymbol{u}_{\star},\label{eq:sample f}
\end{align}
where $\mathbf{W}=\left(\mathbf{B}^{\top}\otimes\mathbb{I}_{T_{\star}}\right)$
is of size $JT_{\star}\times T_{\star}M$ and $\mathbf{f}_{H}=\mathrm{vec}(H\left(\mathbf{F}\right))$
is of size $(JT_{\star})\times1$. Now, Equation (\ref{eq:sample f})
is transformed with a lag polynomial $\Phi_{R}(L)$ in a row-wise
manner. $\Phi_{R}(L)$ has the same functional form as $\Phi_{C}(L)$,
but operates independently on each voxel time series. The first $T_{\star}$
rows will first be transformed, followed by transformation of the
next $T_{\star}$ rows, until all rows have been transformed. The
transformation results in

\[
\tilde{\mathbf{g}}=\mathbf{\tilde{W}}\mathbf{f}_{H}+\boldsymbol{\epsilon},
\]
where $\tilde{\mathbf{W}}$ is of size $JT\times T_{\star}M$, $\tilde{\mathbf{g}}$
and $\boldsymbol{\epsilon}$ are both of size $(JT)\times1$. The
likelihood for $\mathbf{f}_{H}$ is given by

\begin{equation}
\begin{array}{c}
p\left(\tilde{\mathbf{g}}|\mathbf{\tilde{W}},\boldsymbol{\Omega},\mathbf{f}_{H},\boldsymbol{\rho}\right)=\\
\frac{1}{\sqrt{\left(2\pi\right)^{t+k}|\boldsymbol{\Sigma}|}}exp\left(-\frac{1}{2}\left(\tilde{\mathbf{g}}-\mathbf{\tilde{W}}\mathbf{f}_{H}\right)^{\top}\boldsymbol{\Sigma}^{-1}\left(\tilde{\mathbf{g}}-\mathbf{\tilde{W}}\mathbf{f}_{H}\right)\right)
\end{array}\label{eq: GP likelihood}
\end{equation}
Note that $\mathbf{F}$ enters the Gaussian likelihood in a non-linear
way, which means that the full conditional posterior is not available
in closed form. We use elliptical slice sampling \citep{murray2010elliptical}
to sample from the posterior of $\mathbf{F}$. Elliptical slice sampling
is a slice sampling technique which is particularly suitable for Gaussian
process models with non-Gaussian likelihoods.

\section*{Appendix D: Starting values for MCMC\label{App:StartingValues}}

Starting values for the MCMC is obtained in the following fashion:
\begin{itemize}
\item $\mathbf{F}$: the prior mean is used as starting value
\item $\mathbf{Q}$, $\boldsymbol{\sigma}^{2}$ and $\boldsymbol{\rho}$:
Are estimated in a Cochrane-Orcutt estimation procedure. This is done
through iterating estimating the regression parameters ($\mathbf{Q}$
and $\boldsymbol{\sigma}^{2}$) and the time series parameters $\boldsymbol{\rho}$.
Details are given in the Algorithm \ref{alg:Cochrane-Orcutt}.
\end{itemize}
\begin{algorithm}
Iterate:
\begin{enumerate}
\item $\mathbf{Q}$, $\boldsymbol{\sigma}^{2}$: Estimate with Group-wise
ridge regression, using the R-package glmnet \citet{friedman2010regularization}
\item $\boldsymbol{\rho}$: Estimate the autoregressive parameters for the
whole parcel using regularized heteroscedastic regression, by using
the mean of the full conditional posterior of (\ref{eq:rho_full_cond_post}).
\item Stop if difference in mean squared error was less then 0.01 between
two iterations.
\end{enumerate}
\caption{Schematic of the Cochrane-Orcutt estimation procedure.\label{alg:Cochrane-Orcutt}}
\end{algorithm}

\section*{Appendix E: Smooth FIR model for predicted BOLD\label{sec:Appendix FIR}}

The proposed model is also compared with a smooth FIR model. In order
to make the comparison as fair as possible, the FIR model is formulated
as

\begin{equation}
\mathbf{Y}_{\star}=H\left(\mathbf{X}_{FIR}\boldsymbol{h}\right)\mathbf{B}+\mathbf{Z}_{\star}\boldsymbol{\Gamma}+\mathbf{U}_{\star},\label{eq:FIR model}
\end{equation}
where $\mathbf{X}_{FIR}$ is the standard FIR design matrix (where
stimuli are organized column-wise) with $K\times M$ columns. $K$
is the filter length, and $\boldsymbol{h}$ is the filters for all
stimuli stacked in one vector of size $KM\times1$. $\boldsymbol{h}$
has independent and identical GP priors for each stimuli. Since $\boldsymbol{h}$
contains much fewer parameters than $\mathbf{F}$, the scale factor
of the kernel is specified to a higher value compared to $\mathbf{F}$,
in order to increase flexibility. The filters in $\boldsymbol{h}$
is estimated using elliptical slice sampling, in the same way as $\mathbf{F}$
is estimated in our model. The inference for the other parameters
is not changed. The prior mean function for $\boldsymbol{h}$ was
specified to the common double gamma HRF, and the kernel hyperparameters
were specified to $l=3$ and $\omega=\sqrt{0.5}\approx0.707$. The
endpoints of the filter were constrained to the corresponding prior
mean values.
\end{document}